\documentclass[journal=jacsat,manuscript=article]{achemso}

\usepackage{amsmath}
\usepackage{amsmath,amssymb}
\usepackage{graphicx}
\usepackage{caption}
\usepackage{color}
\usepackage{dcolumn}
\usepackage{bm}
\usepackage{float}
\usepackage{subfig}
\usepackage{soul}
\usepackage{comment}

\usepackage{xr}
\usepackage{rotating}

\SectionNumbersOn \author{Sugata Goswami, Juan Carlos San Vicente
  Veliz, Meenu Upadhyay} \affiliation{Department of Chemistry,
  University of Basel, Klingelbergstrasse 80, CH-4056 Basel,
  Switzerland}

\author{Raymond J. Bemish} \affiliation{Air Force Research Laboratory,
  Space Vehicles Directorate, Kirtland AFB, New Mexico 87117, USA}

\author{Markus Meuwly} {\affiliation{Department of Chemistry,
    University of Basel, Klingelbergstrasse 80, CH-4056 Basel,
    Switzerland} \altaffiliation{Department of Chemistry, Brown
    University, RI, USA} \email{m.meuwly@unibas.ch}

\title{Quantum and Quasi-classical Dynamics of the C($^{3}$P) +
  O$_{2}$($^3\Sigma_{g}^{-}$) $\rightarrow$ CO($^{1}\Sigma^{+}$)+
  O($^{1}$D) Reaction on Its Electronic Ground State}

\begin{document}

\date{\today}

\begin{abstract}
The dynamics of the C($^{3}$P) + O$_{2}$($^3\Sigma_{g}^{-}$)
$\rightarrow$ CO($^{1}\Sigma^{+}$)+ O($^{1}$D) reaction on its
electronic ground state is investigated by using time-dependent wave
packet propagation (TDWP) and quasi-classical trajectory (QCT)
simulations. For the moderate collision energies considered ($E_{\rm
  c} = 0.001$ to 0.4 eV, corresponding to a range from 10 K to 4600 K)
the total reaction probabilities from the two different treatments of
the nuclear dynamics agree very favourably. The undulations present in
$P(E)$ from the quantum mechanical treatment can be related to
stabilization of the intermediate CO$_2$ complex with lifetimes of on
the 0.05 ps time scale. This is also confirmed from direct analysis of
the QCT trajectories. Product diatom vibrational and rotational level
resolved state-to-state reaction probabilities from TDWP and QCT
simulations also agree well except for the highest product vibrational
states $(v' \geq 15)$ and for the lowest product rotational states
$(j' \leq 10)$. Opening of the product vibrational level CO$(v' = 17)$
requires $\sim 0.2$ eV from QCT and TDWP simulations with O$_2$($j=0$)
and decreases to 0.04 eV if all initial rotational states are included
in the QCT analysis, compared with $E_{\rm c} > 0.04$ eV obtained from
experiments. It is thus concluded that QCT simulations are suitable
for investigating and realistically describe the C($^{3}$P) +
O$_{2}$($^3\Sigma_{g}^{-}$) $\rightarrow$ CO($^{1}\Sigma^{+}$)+
O($^{1}$D) reaction down to low collision energies when compared with
results from a quantum mechanical treatment using TDWPs.
\end{abstract}

\maketitle

\section{Introduction}
Reactions involving carbon and oxygen atoms play a vital role in the
atmosphere, in combustion, and in hypersonic flow.\cite{Sharma2010}
Among all, the C($^{3}$P) + O$_{2}$($^3\Sigma_{\rm g}^{-}$)
$\longleftrightarrow$ CO($^{1}\Sigma^{+}$)+ O($^{1}$D)/O($^{3}$P)
reaction involving the CO$_2$ intermediate and several electronic
states is particularly important and has, therefore, been investigated
both, experimentally and through
computations.\cite{husain:1975,BERGEAT1999,
  Geppert2000,becker:1988,dorthe:1991, chastaing:1999, chastaing:2000,
  Hanson:1991,thrush:1973, dubrin:1964, Xantheas1994, Jasper2013,
  troe:1975,Braunstein2000, shatalov:2000, Center1973, Kelley1977,
  Davidson1978,Tully:1975,VelizPCCP2021} The electronic ground and
excited state potential energy surfaces (PESs) were studied
computationally in detail at different levels of
theory\cite{Xantheas1994,winter:1973,greben:2012,greben:2013,greben:2013.2,Schmidt2013,hsien:2013,VelizPCCP2021}
and various dynamics simulations were also carried out for this
system.\cite{Jasper2013,Braunstein2000,murrell:1977,Brunsvold2008,VelizPCCP2021}\\

\noindent
One of the latest investigations for the entire C($^{3}$P) +
O$_{2}$($^3\Sigma_{g}^{-}$) $\longleftrightarrow$ CO$_{2}$
$\longleftrightarrow$ CO($^{1}\Sigma^{+}$)+ O($^{1}$D)/O($^{3}$P)
reaction was carried out using quasi-classical trajectory (QCT)
simulations with kernel-represented PESs based on extensive multi
reference configuration interaction (MRCI)
calculations.\cite{VelizPCCP2021} The focus was more on the
high-temperature regime of the reaction also because experimental and
computational data from shock tube experiments for the C + O$_{2}$
$\rightarrow$ CO + O reaction between 1500-4200 K was
available.\cite{Hanson:1991,Jasper2013} The latest QCT simulations
reported the forward and backward thermal rates for the lowest five
electronic states together with vibrational relaxation
times.\cite{VelizPCCP2021} However, the state-to-state dynamics was
not considered and the question remains whether a classical framework
is suitable when compared with results from a quantum mechanical
treatment of the nuclear dynamics.\\

\noindent
One recurrent theme for atom + diatom reactions concerns the range of
applicability of quasi-classical-based dynamics approaches for
computing thermal rates and final state distributions. Cross sections
and thermal rates are averaged over initial and/or final
ro-vibrational resolved state-to-state information. Thus, although
more or less heavily averaged quantities may favourably agree between
different approaches, it is possible that such agreement arises from
the averaging process. Hence, it is also relevant to compare
properties at the state-to-state level. All such quantities are
essential as input to more coarse-grained investigations of reaction
networks as they appear in hypersonics, combustion, atmospheric and
astrophysical chemistry.\cite{MM.rev:2020} Because QCT simulations are
computationally more efficient, they are often used instead of and
also along with more time-consuming quantum nuclear dynamics
simulations.\cite{zanchet2009study,goswami2014time,jorfi2009quasi,jorfi2009quasiclassical,jorfi2009statistical}\\

\noindent
Quantum mechanical and QCT approaches were used to evaluate reaction
probabilities and rates for the C+OH reaction and were found to be in
fairly good agreement.\cite{jorfi2010quantum,jorfi2011state,rao:2013}
For the S+OH reaction quantum mechanical and QCT simulations produced
excellent agreement for cross sections but only fair agreement for the
total reaction probabilities and thermal
rates.\cite{jorfi2011quasi,goswami2014time} Comprehensive QCT and QM
dynamics investigations for the Br + H$_{2}$ and O + HCl reactions
reported favourable agreement between QM and QCT results for
properties such as reaction probabilities, integral and differential
cross sections.\cite{panda2012state,bargueno2011energy} Similar to the
S+OH reaction, for N + H$_{2}$ results from QM and QCT simulations
agree favourably for cross sections but not so well for reaction
probabilities.\cite{hankel2012quantum}\\

\noindent
For the S+OH reaction it has been specifically reported that for
reactive collisions a more general relationship between the
mechanistic details of the dynamics and the ensuing rates can be
difficult to obtain.\cite{goswami2014time} Nonetheless, it is
generally believed that the dynamical features of an exoergic and
barrierless reaction primarily depend on the masses of the
participating atoms, the exoergicity and the topographical details of
the underlying
PES. \cite{GoswamiMolPhys2017,goswami2014time,goswami2018theoretical}
To better understand the origins of the observed dynamics systematic
investigations have been carried out for different
systems.\cite{polanyi1972some,GoswamiMolPhys2017,goswami2018theoretical,rao:2013,jiang2013relative,polanyi1987some}
As an example, for the C + OH reactive
collisions\cite{GoswamiMolPhys2017,goswami2018theoretical,rao:2013} it
was found that reagent vibrational excitation decreased reactivity on
the first electronically excited PES and enhances reactivity on the
second excited state. The excess vibrational energy is transferred
into product translational energy for the first excited PES whereas
for the second excited state it is transferred into product vibration
and rotation. These dynamical effects on the final states are caused
by the topology of the underlying PESs which have no barrier to
products for the first electronically excited state but involve a late
barrier for the second excited state.\\

\noindent
The reaction C($^{3}$P) + O$_{2}$($^3\Sigma_{g}^{-}$) $\rightarrow$
CO($^{1}\Sigma^{+}$)+ O($^{1}$D)/O($^{3}$P) using its five lowest
lying states ($^{1}$A${'}$, $^{3}$A${'}$, $^{3}$A${''}$, $^{1}$A${''}$
and (2)$^{1}$A${'}$) is a particularly suitable system for such an
investigation due to the availability of high-quality,
full-dimensional PESs which were recently validated vis-a-vis
experiment.\cite{VelizPCCP2021} The exoergicity of the three singlet
and two triplet states are $\sim3.8$ eV and $\sim5.9$ eV,
respectively.\cite{VelizPCCP2021} Therefore, the difference in the
dynamical attributes on the singlet and triplet states is expected to
arise solely from the topographical details of the underlying
surfaces. \cite{goswami2014time}\\

\noindent
Quantum mechanics-based dynamics approaches can become computationally
expensive for barrierless exoergic reactions with deep
wells.\cite{GoswamiMolPhys2017,goswami2014time,goswami2018theoretical,rao:2013,bulut2011accurate,lin2008accurate}
Thus, QCT approaches are an attractive alternative, at least at the
qualitative level. In the present work the initial state-selected and
state-to-state reaction probabilities for the C($^{3}$P) +
O$_{2}$($^3\Sigma_{g}^{-}$) $\rightarrow$ CO($^{1}\Sigma^{+}$)+
O($^{1}$D) reaction on the electronic ground ($^{1}$A${'}$) state are
determined for low to moderate collision energies ($E_{\rm c} \leq
0.4$ eV) by TDWP and QCT approaches. Total reaction probabilities and
final state distributions are determined for a range of collision
energies and allow to compare the two dynamics methods in terms of
their findings and with experiment. The work is organized as
follows. First the methods used are described. Then, the convergence
of the TDWP simulations is assessed, followed by results for the total
reaction probabilities and the product vibrational resolved reaction
probabilities from TDWP and QCT simulations. Next, product vibrational
and rotational state distributions are determined and
compared. Finally, the computational results are discussed in the
context of experiments and in a broader sense.\\

\section{Methods}

\subsection{Time Dependent Wavepacket Calculations}
\begin{figure}
\includegraphics*[scale=0.4]{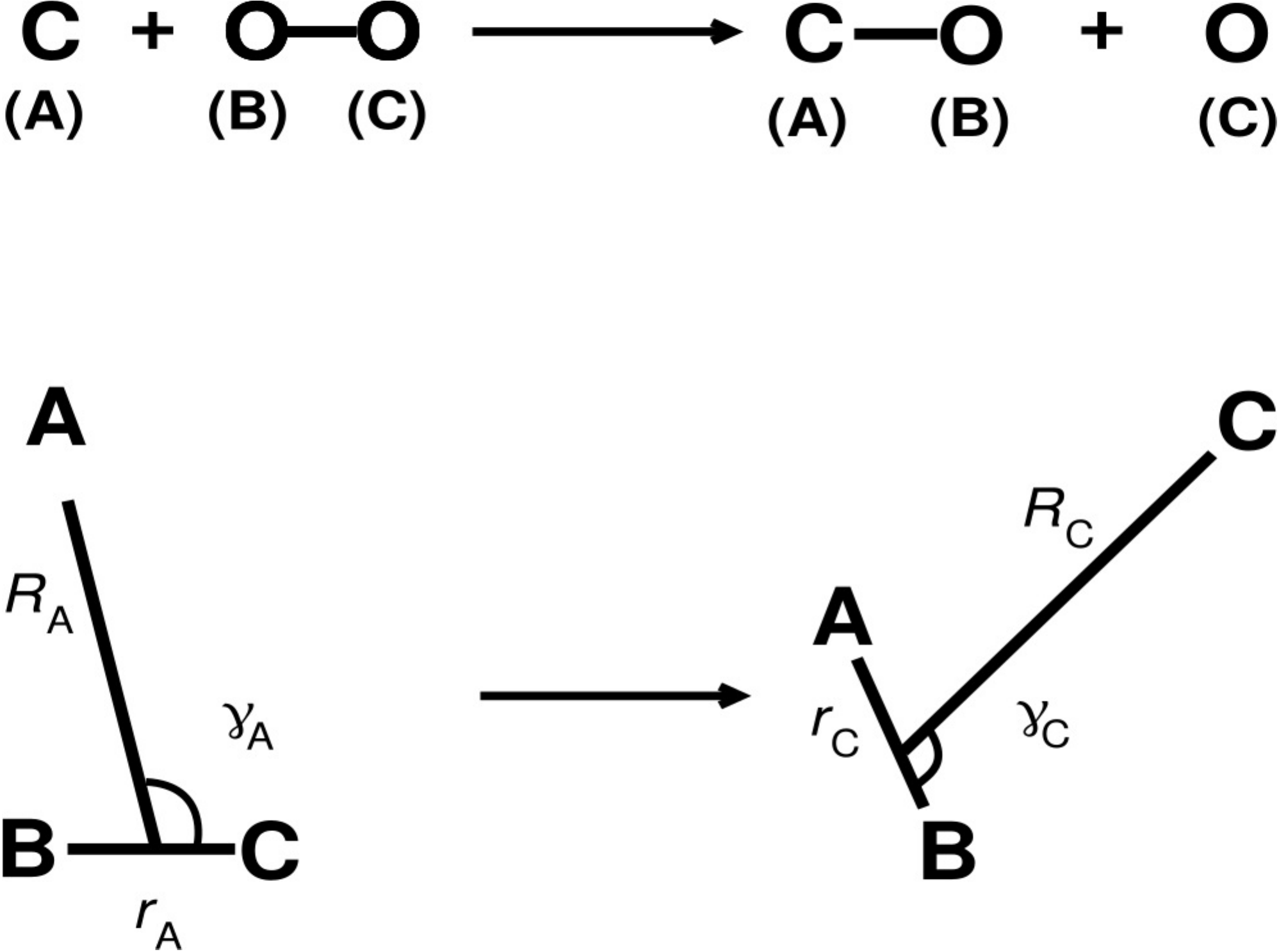}
\caption{Reactant (left) and product (right) collision geometries in
  body-fixed Jacobi coordinates for the present TDWP simulations.}
\label{fig:fig1}
\end{figure}

State-to-state quantum dynamics was investigated using the modified
DIFFREALWAVE code which is based on propagating real wave packets
(RWP).\cite{hankel2008diffrealwave,gray1998quantum} Here only a
summary of the method employed is given as the theory has been
described extensively
elsewhere.\cite{gray1998quantum,hankel2008diffrealwave,hankel2006state,goswami2018theoretical,goswami2020effect}
The RWP approach is advantageous from a computational perspective as
the formalism requires only the real part of the wave packet (WP) to
be propagated for obtaining the state-to-state $S$-matrix elements.
An initial WP, $q_{A}$($R_{\rm A}, r_{\rm A}, \gamma_{A}, t=0$) is
prepared in the asymptotic reactant channel with the geometry
described in a Jacobi coordinate system ($R_{\rm A}$: distance of atom
A from the center-of-mass of BC, $r_{\rm A}$: BC internuclear distance
and $\gamma_{\rm A}$: approach angle of A to the center-of-mass of
BC), see Figure \ref{fig:fig1}.  The real part of such a WP in this
coordinate system takes the form
\cite{hankel2006state,goswami2018theoretical,goswami2020effect}
\begin{equation}
 q_{\rm A} (R_{A}, r_{A}, \gamma_{A}, t=0) = \frac{{\rm \sin}[\alpha (R_{\rm A}-R_{0})]}{R_{\rm A}-R_{\rm 0}} {\rm \cos}[k_{\rm 0}(R_{\rm A}-R_{\rm 0})] 
 e^{-\beta_{s}(R_{\rm A}-R_{\rm 0})^{2}} \phi^{BC}_{vj}(r_{\rm A})P^{\Omega}_{j}({\rm \cos} \gamma_{\rm A})
 \label{eq:eq1}
\end{equation}
The parameters $\alpha$ and $\beta_{s}$ define the width and the
smoothness of the WP, respectively,\cite{hankel2003sinc} $k_{\rm 0}$
is the initial momentum of the WP centered at $R_{\rm 0}$, $\phi^{\rm
  BC}_{vj}$ and $P^{\Omega}_{j}$(cos$\gamma_{a}$) represent the
ro-vibrational eigenfunction of BC and the associated Legendre
polynomials, respectively, and $v$ and $j$ are the initial vibrational
and rotational levels of the reactant diatom. \\

\noindent
The initial WP (Eq. (\ref{eq:eq1})) is subsequently transformed to
product state Jacobi coordinates ($R_{\rm C}$, $r_{\rm C}$,
$\gamma_{\rm C}$) with C as atom and AB as diatom, see right-bottom
part of Figure \ref{fig:fig1}, according to\cite{hankel2006state}
\begin{equation}
 q^{J\Omega{'}}_{\rm C} (R_{\rm C}, r_{\rm C}, \gamma_{\rm C}, t=0) =
 N\sqrt{w_{\rm C}}q_{\rm A} (R_{\rm A}, r_{\rm A}, \gamma_{\rm A},
 t=0) \frac{R_{\rm C}r_{\rm C}}{R_{\rm A}r_{\rm A}}d^{J}_{\Omega
   \Omega{'}}(\beta).
 \label{eq:eq2}  
\end{equation}
Here $N$, $d^{J}_{\Omega \Omega{'}}(\beta)$, $\beta$ and $\Omega{'}$
are normalization, the reduced Wigner matrix, the angle between the
two vectors, $\textbf R_{\rm C}$ and $\textbf R_{\rm A}$, and the
projection of $J$ on the body-fixed $Z$-axis of the product Jacobi
coordinates, respectively.\\

\noindent
The action of the nuclear Hamiltonian $\hat{H}_{\rm nuc}$, formulated
in product Jacobi coordinates, $\hat{H}_{\rm nuc}$, on the WP
is\cite{hankel2006state,offer1994time}
\begin{eqnarray}
\hat{H}_{\rm nuc}q^{J\Omega'}_{\rm C}(t)&=&
\left[-\frac{1}{2\mu_{R}^{AB,C}}\frac{\partial ^{2}}{\partial
    R_{\rm C}^{2}}-\frac{1}{2\mu_{r}^{AB}}\frac{\partial^{2}}{\partial
    r_{\rm C}^{2}}\right]q^{J\Omega'}_{\rm C}(t) \nonumber
\\ &-&\left(\frac{1}{2\mu_{R}^{AB,C}R_{\rm C}^{2}}+\frac{1}{2\mu_{r}^{AB}r_{\rm C}^{2}}\right)\left[\frac{1}{{\rm
      \sin}\gamma_{\rm C}}\frac{\partial}{\partial\gamma_{\rm C}}{\rm
    \sin}\gamma_{\rm C}\frac{\partial}{\partial\gamma_{\rm C}}-\frac{\Omega'^{2}}{{\rm
      \sin}^{2}\gamma_{\rm C}}\right] q^{J\Omega'}_{\rm C}(t)\nonumber
\\ &+&\hat{V}(R_{\rm C}, r_{\rm C}, \gamma_{\rm C})q^{J\Omega'}_{\rm C}(t) \nonumber
\\ &+&\left(\frac{1}{2\mu_{R}^{AB,C}R_{\rm C}^{2}}\right)(J(J+1)-2\Omega'^{2})q^{J\Omega'}_{\rm C}(t)
\nonumber
\\ &-&\frac{C^{+}_{J\Omega'}}{2\mu_{R}^{AB,C}R_{\rm C}^{2}}\left[\frac{\partial}{\partial\gamma_{\rm C}}-\Omega'
  \cot\gamma_{\rm C}\right]q^{J(\Omega'+1)}_{\rm C}(t) \nonumber
\\ &-&\frac{C^{-}_{J\Omega'}}{2\mu_{R}^{AB,C}R_{\rm C}^{2}}\left[-\frac{\partial}{\partial\gamma_{\rm C}}-\Omega'
  \cot\gamma_{\rm C}\right]q^{J(\Omega'-1)}_{\rm C}(t).
 \label{eq:eq3}
\end{eqnarray}
Here and in what follows, the coordinate dependence of the WPs,
$q^{J\Omega'}_{\rm C}(t)$, $q^{J(\Omega'+1)}_{\rm C}(t)$ and
$q^{J(\Omega'-1)}_{\rm C}(t)$ is omitted for clarity. The terms
$\mu^{AB,C}_{R}$(=$\frac{m_{\rm C}(m_{\rm A}+m_{\rm B})}{(m_{\rm
    A}+m_{\rm B}+m_{\rm C})}$) and $\mu_{r}^{AB}$(=$\frac{m_{\rm
    A}m_{\rm B}}{(m_{\rm A}+m_{\rm B})}$) represent the three body and
product diatom reduced masses, respectively. The action of the radial
kinetic energy part of the Hamiltonian is carried out using fast
Fourier transformation techniques.\cite{kosloff1983fourier} For the
action of the angular kinetic energy part a discrete variable
representation based on Gauss-Legendre quadrature is
used.\cite{light1985generalized,lill1982discrete,hamilton1986distributed}
The effect of the potential energy operator, $\hat{V}$($R_{\rm C},
r_{\rm C}, \gamma_{\rm C}$), on the WP is multiplicative on the
coordinate grid and the last two terms of Eq.
(\ref{eq:eq3}) denote the Coriolis coupling and lead
to mixing of the WP, $q^{J\Omega{'}}_{{\rm C}}$ with other WPs,
$q^{J(\Omega{'}+1)}_{\rm C}$ and $q^{J(\Omega{'}-1)}_{\rm C}$ as shown
in the Eq. (\ref{eq:eq3}).\\

\noindent
The propagated WP at the first time step, $\tau_{1}$, is evaluated
according to
\begin{equation}
 q^{J\Omega{'}}_{\rm C}(\tau_{1}) = \hat{H}_{\rm nuc, s}q^{J\Omega{'}}_{\rm C}(0) 
                                 -\sqrt{1-\hat{H}^{2}_{\rm nuc, s}}p^{J\Omega{'}}_{\rm C}(0), 
\label{eq:eq4}                                 
\end{equation}
where $p^{J\Omega{'}}_{\rm C}$ denotes the imaginary part of the
initial WP (see Eq. (\ref{eq:eq1}) and
(\ref{eq:eq2})). For the
subsequent time propagation a three term recursion is used
\begin{equation}
 q^{J\Omega{'}}_{\rm C}(t+\tau) = -q^{J\Omega{'}}_{\rm C}(t-\tau) +
 2\hat{H}_{nuc, s}q^{J\Omega{'}}_{\rm C}(t),
\label{eq:eq5} 
\end{equation}
where $\tau$ is the discrete time step. In
Eq. (\ref{eq:eq4}) and
(\ref{eq:eq5}), $\hat{H}_{nuc,s}$
represents the scaled and shifted Hamiltonian whose minimum and
maximum eigenvalues lie between $-$1 and $+$1.\cite{gray1998quantum}\\

\noindent 
As is customary for TDWP simulations, spurious reflections of the WP
components from the finite sized grid edges at longer propagation time
need to be damped. This is done by multiplying the wavepacket with a
double exponential damping function\cite{hankel2006state} $a(y)$ along
coordinates $y = {R_{c},r_{c}}$
\begin{equation}
  a(y) = {\rm \exp}[-c_{abs}{\rm \exp}(-2(y_{max}-y_{abs})/(y-y_{abs}))]
 \label{eq:eq6} 
\end{equation}
according to $q^{J\Omega{'}}_{\rm C}(t) \cdot a(R_{\rm C}) \cdot
a(r_{\rm C})$, where $c_{abs}$ and $y_{abs}$ represent the strength
and the starting point of the absorption, respectively. The absorption
is applicable for $y>y_{abs}$ and the value of the function is equal
to 1 elsewhere.\\

\noindent
The propagated WP is projected on ro-vibrational levels of the product
diatom at the product asymptote ($R = R_{\infty}$) to obtain the
time-dependent coefficients \cite{hankel2006state}
\begin{equation}
 C^{J}_{v,j,\Omega \rightarrow v{'},j{'},\Omega{'}}(t) = \int
 \phi^{*AB}_{v{'},j{'}}(r_{\rm C}, \gamma_{\rm C})q^{J\Omega{'}}_{\rm
   C}(R_{\rm C}=R_{\infty},r_{\rm C},\gamma_{\rm C},t)dr_{\rm C}{\rm
   \sin}(\gamma_{\rm C})d\gamma_{\rm C}.
\label{eq:eq7}   
\end{equation}
The term, $\phi^{AB}_{v{'},j{'}}$ in Eq. (\ref{eq:eq7}) represents the
$(v{'},j{'})$ state of the product diatom AB. The time-dependent
coefficients $C^{J}_{v,j,\Omega \rightarrow v{'},j{'},\Omega{'}}(t)$
are Fourier transformed to obtain energy dependent coefficients,
$A^{J}_{v,j,\Omega \rightarrow v{'},j{'},\Omega{'}}(E)$ in the body
fixed frame and then are transformed to the space fixed frame from
which the scattering matrix ($S$-matrix) elements, $S^{J}_{v,j,l
  \rightarrow v',j'l'}(E)$ are obtained. These $S$-matrix elements are
finally transferred back to the body-fixed frame to yield
$S^{J}_{v,j,\Omega \rightarrow v{'},j{'}\Omega{'}}(E)$ from which
energy dependent state-to-state reaction probabilities are obtained
according to
\begin{equation}
 P^{J}_{v,j,\Omega \rightarrow v{'},j{'},\Omega{'}}(E) =
 |S^{J}_{v,j,\Omega \rightarrow v{'},j{'},\Omega{'}}(E)|^{2}
\label{eq:eq8} 
\end{equation}
Product internal level resolved probabilities are calculated by
summing up the probabilities of Eq. \ref{eq:eq8} over
relevant quantum numbers. Summation over all three quantum numbers,
$(\Omega{'}, j{'}, v{'})$ yields the energy resolved total reaction
probabilities. For further details, see
Ref.\citenum{hankel2006state}.\\

\subsection{Quasi Classical Trajectory Simulations}
The QCT simulations in the present investigation were carried
following earlier work\cite{VelizPCCP2021} and based on established
procedures.\cite{hen11,kon16:4731,MM.cno:2018,tru79} Therefore, only
specific technical aspects are briefly summarized here.\\

\noindent
Hamilton's equations of motion were solved using a fourth order
Runge-Kutta method. The initial conditions for initiating the
trajectories were sampled by using standard Monte Carlo
methods\cite{tru79} and the ro-vibrational levels of reactant and
product diatoms are calculated using semiclassical quantization. The
simulations are run at several collision energies with a time step of
$\Delta t=0.05$ fs which guarantees the conservation of total energy
and angular momentum. At each collision energy, $5 \times 10^{5}$
trajectories are run for converged results, except at the two lowest
energies, 0.0055 and 0.001022 eV for which $10^{6}$ trajectories are
run for convergence of the total reaction probability for
O$_{2}$($v=0$, $j=0$). It is found that the difference between the
total reaction probabilities at these two energies obtained from $5
\times 10^{5}$ and $10^{6}$ trajectories appears at the fourth decimal
place and therefore all remaining calculations are carried out with $5
\times 10^{5}$ trajectory simulations. As the associated quantum
numbers are real-valued, their necessary assignment to integer values
was made either by using Gaussian binning
(GB)\cite{bon97:183,bon04:106,kon16:4731} which centers Gaussian
weights around the integer values and has a full width at half maximum
of 0.1 or from Histogram binning (HB) which rounds to the nearest
integer.\\

\noindent
The reaction probability at collision energy $E_{\rm col}$ is obtained
as
\begin{equation}
    P_{\rm r}(E_{\rm col})=\frac{N_{r}(E_{\rm col})}{N_{\rm tot}(E_{\rm col})},
    \label{eq:eq9} 
\end{equation} \\
where $N_{\rm r}(E_{\rm col})$ is the number of reactive trajectories
and $N_{\rm tot}(E_{\rm col})$ is the total number of trajectories at
a give $E_{\rm col}$.\\

\section{Results and Discussion}
\subsection{Convergence of the TDWP approach}
The TDWP approach is a grid-based method and the results depend on the
parameters characterizing the wavepacket and the underlying grids. In
a first step, convergence of the total reaction probability was sought
for each parameter. For a particular parameter considered (see Table
\ref{tab:tab1}) its value was changed until the total reaction
probability remained unchanged. The convergence runs were initialized
from arbitrary guesses for the parameters and the wave packet
propagation was carried out for fewer time steps (25000 in the present
case) than for the final production runs. If convergence was not
achieved, the WP was propagated for longer times.\\

\begin{table}
\caption{Details of the numerical parameters related to the coordinate
  grid, properties of initial wave packet and the damping/ absorption
  function used in present TDWP investigation of the C + O$_{2}$
  ($v=0-2$, $j=0$) $\rightarrow$ CO + O reaction on its electronic
  ground ($^{1}$A$'$) state.}
\vspace*{0.1in}
\begin{tabular}{lll}
\hline \hline
Parameter & Value & Description\\
\hline
$N_R/N_r/N_{\gamma}$  & 499/419/240 & Number of grid points \\
                                     &&along the three product Jacobi coordinates, \\
                                     &&$R_{\rm C}$, $r_{\rm C}$ and $\gamma _{\rm C}$. \\
$R_{min}/R_{max}$ ($a_{0}$) & 2.2/22.0 & Extension of the grid along $R_{\rm C}$ \\
$r_{min}/r_{max}$ ($a_{0}$)& 1.5/17.9902 & Extension of the grid along $r_{\rm C}$ \\
$R_{\infty}$ ($a_{0}$) & 12.5 & Location of the dividing surface\\
                           && in the product channel \\
$R_{abs}/r_{abs}$ ($a_{0}$) & 14.0/14.75 & Starting point of the absorption \\
                                     && function along $R_{\rm C}$ and $r_{\rm C}$\\
$C_{abs}/c_{abs}$                 & 30.0/30.0 & Strength of the absorption along $R_{\rm C}$ and $r_{\rm C}$ \\
$R_0$ ($a_{0}$) & 13.2 & Center of the initial wave packet in \\
                         && reactant Jacobi coordinate \\
$E_{trans}$ (eV) & 0.115 & Initial translational energy in eV\\
$\alpha$ & 12.0 & Width of the initial wave packet \\
$\beta_{s}$ & 0.5 & Smoothing of the initial wave packet\\
\hline \hline
\end{tabular}
\label{tab:tab1}
\end{table}

\noindent
Convergence runs for all parameters were carried out for $J=0$ and for
reactant O$_{2}(v=0, j=0)$. The converged parameters are summarized in
Table \ref{tab:tab1} and demonstrate that extensive grids along the
three product Jacobi coordinates are required resulting in
computationally expensive calculations. The requirement for such large
grids is the result of the heavy masses of the interacting atoms and
the presence of the deep potential well.\cite{VelizPCCP2021} The large
number of grid points required along the angular Jacobi coordinate
reflects the pronounced angular anisotropy of the underlying PES. \\

\begin{figure}
 \begin{center}
  \includegraphics*[scale=0.70]{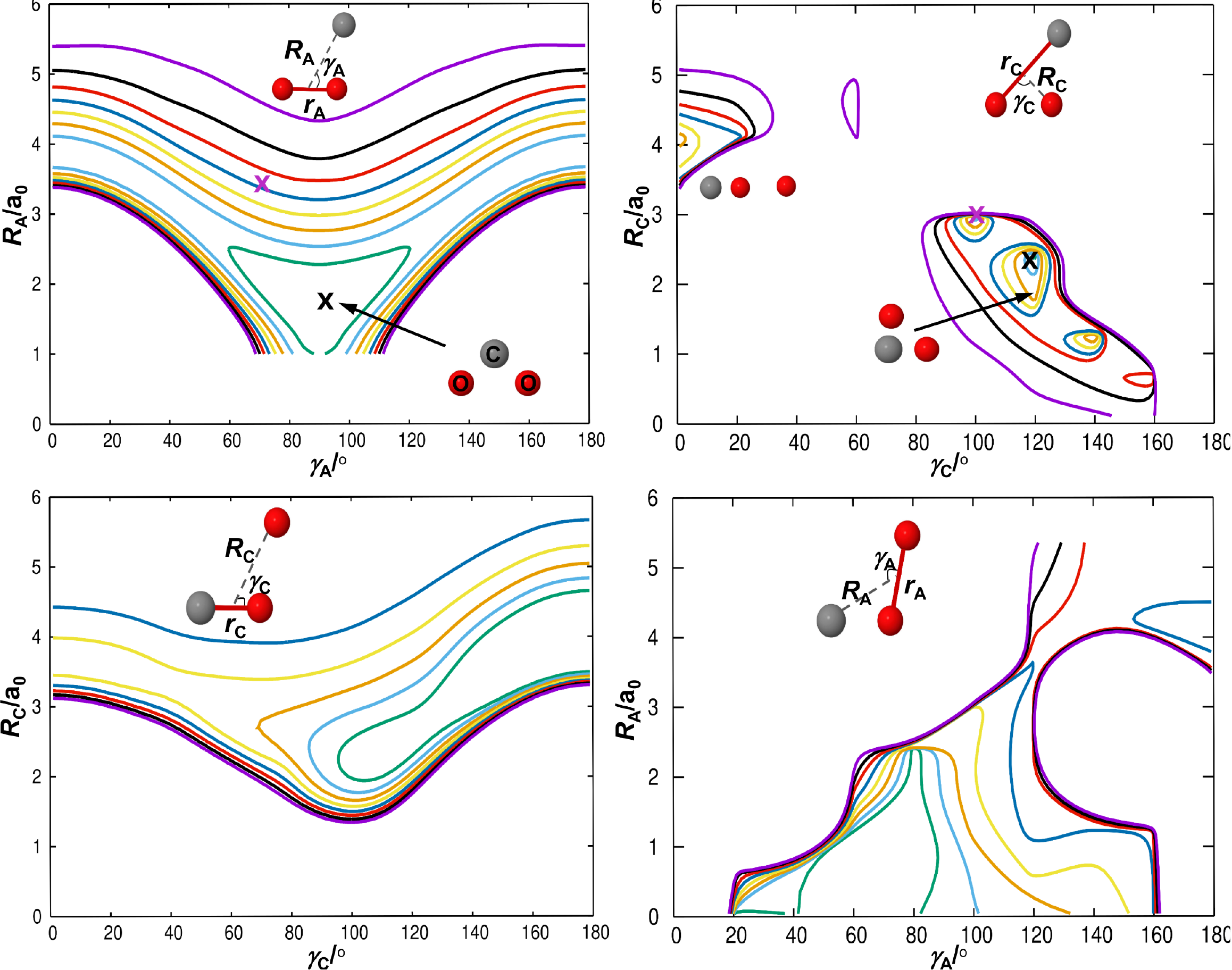}
  \caption{Two-dimensional contour diagrams of the $^{1}$A${'}$ PES
    for CO$_{2}$ for both Jacobi coordinate systems (reactant: $R_{\rm
      A}, \gamma_{\rm A}$) and (product: $R_{\rm C}, \gamma_{\rm
      C}$). Top: $r_{\rm OO} = \sim 3.4$ a$_0$ for reactant (left) and
    product (right) coordinates. Bottom: $r_{\rm CO} = 3.0$ a$_0$ for
    product (left) and reactant (right) coordinates. Contours are
    drawn at $-2.0$ eV (violet), $-3.0$ eV (black), $-4.0$ eV (red),
    $-5.0$ eV (blue), $-6.0$ eV (yellow), $-7.0$ eV (orange), $-8.0$
    eV (cyan), $-9.0$ eV (green) which are with respect to the total
    atomization energy C($^{3}$P) + O($^{3}$P) + O($^{3}$P). In each
    panel, a representative structure for CO$_2$ is shown in terms of
    the respective Jacobi coordinates. Structures for CO$_2$ (carbon
    atom grey and oxygen atoms red spheres) near the minima are
    schematically shown and their position on the PES is indicated
    with arrows. The black and purple crosses indicate equivalent
    geometries in the two different Jacobi coordinates.}
 \label{fig:fig2}   
 \end{center}
\end{figure}

\noindent
Two-dimensional (2-D) contours of the underlying PES are shown in
Figure \ref{fig:fig2}. The top row reports the PES in reactant Jacobi
coordinates for $R_{\rm OO} = r_{\rm A} = 3.40$ a$_0$, i.e. a
stretched geometry of the reactant diatom before C-insertion occurs to
form O-C-O (minimum at $90^\circ$). Because after formation of the
triatom the system only spends little time in this state (typically $<
0.5$ ps, see Discussion section) to move towards the product, it is
also of interest to provide a representation of the reactant PES in
terms of product Jacobi coordinates ($R_{\rm C}, \gamma_{\rm C}$)
which is shown in the top right panel in Figure
\ref{fig:fig2}. Comparison of the two representations clarifies that
spatial symmetry is lost for the product state coordinates and the
anisotropy of the PES differs considerably. Similarly, the bottom row
reports the PES in product state coordinates for $R_{\rm CO} = r_{\rm
  C} = 3.0$ a$_0$, i.e. for the system on its way to form products
(left panel) and the same PES represented in reactant state
coordinates ($R_{\rm A}, \gamma_{\rm A}$) in the right hand lower
panel of Figure \ref{fig:fig2}. In addition, representative structures
for CO$_{2}$ are reported and their location on the PES are indicated
by arrows. Two geometries (see black and purple crosses in the top row
panels) are chosen to show their positions in the reactant and
equivalent product Jacobi coordinates.\\

\noindent
Because converging the wavepacket with O$_{2}(v=0, j=0)$ was
laborious, no reoptimization was carried out for O$_{2}(v=1, j=0)$ and
O$_{2}(v=2, j=0)$. However, the number of product vibrational levels
was suitably increased to $v'=19$ and $v'=20$ with the reactant in its
two vibrationally excited states. The WP was propagated for up to
70000 time steps (corresponding to $\sim 1.4$ ps) in order to obtain
converged reaction probabilities for $J=0$ and the potential and/or
centrifugal cut-off was $0.57 E_{\rm h}$.\\

\subsection{Total Reaction Probabilities}
\begin{figure}
 \begin{center}
  \includegraphics*[scale=0.7]{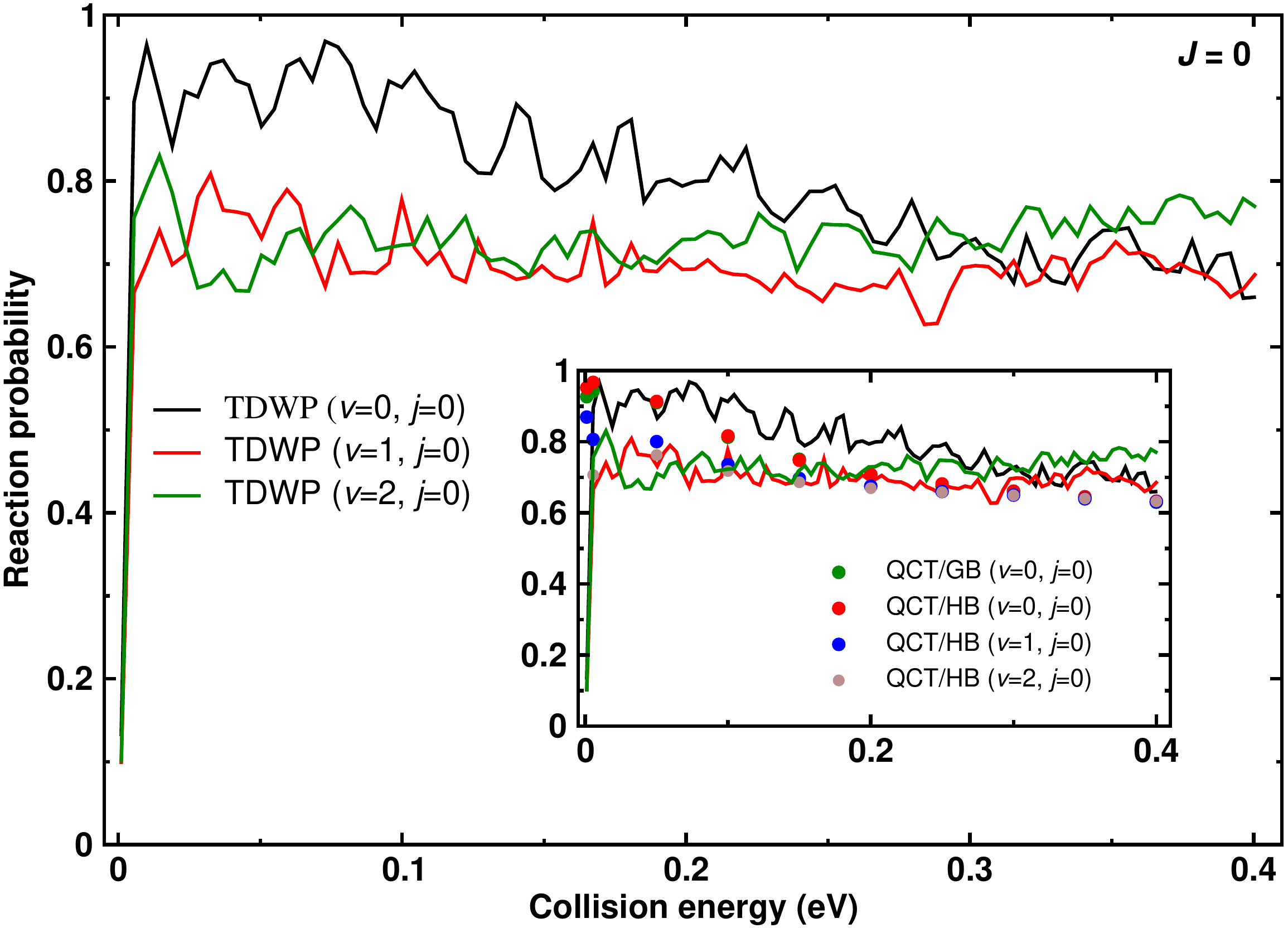}
  \caption{Initial state-selected and energy resolved total reaction
    probabilities of the C($^{3}$P) + O$_{2}$($^3\Sigma_{g}^{-}$,
    $v=0-2$, $j=0$) $\rightarrow$ CO($^{1}\Sigma^{+}$, $\sum v{'}$,
    $\sum j{'}$) + O($^{1}$D) reaction on its electronic ground state
    as a function of collision energy for $J=0$. The TDWP and QCT
    probabilities are shown by solid lines and circles, respectively. The
    red and green circles represent the QCT results obtained by using
    Histogram and Gaussian binning, respectively, for O$_{2}$($v=0$,
    $j=0$). These probabilities at the two lowest collision energies
    are calculated by running 10$^{6}$ trajectories whereas $5 \times
    10^{5}$ trajectories are used for the other energies. The QCT
    probabilities for O$_{2}$($v=1-2$, $j=0$) are shown in the inset
    by blue and brown solid circles.}
   \label{fig:fig3}
 \end{center}
\end{figure}

Initial state-selected and energy resolved total reaction
probabilities of the C($^{3}$P) + O$_{2}$($^3\Sigma_{g}^{-}$, $v=0-2$,
$j=0$) $\rightarrow$ CO($^{1}\Sigma^{+}$, $\sum v{'}$, $\sum j{'}$)+
O($^{1}$D) reaction on its electronic ground state ($^{1}$A$'$) are
shown as function of collision energy in Figure \ref{fig:fig3}. These
probabilities are calculated for total angular momentum $J=0$. The
inset compares the TDWP results (solid lines) with the QCT results
(solid circles), respectively.\\

\noindent
Figure \ref{fig:fig3} shows that starting at low collision energy
($E_{\rm col} = 0.001$ eV $\approx 10$ K) the TDWP probability for
O$_{2}$($v=0$, $j=0$) rises sharply and oscillates around 0.95 up to
$E_{\rm col} \sim 0.1$ eV. Afterwards, the total reaction probability
decreases slowly with undulations and reaches $P \sim 0.7$ at $E_{\rm
  col} \sim 0.4$ eV. The non-zero probability for $E_{\rm col} \sim
10^{-3}$ eV (Figure \ref{fig:fig3}) reflects the barrierless nature of
this exoergic reaction.\cite{VelizPCCP2021} The barrierless and
exoergic features of this reaction along with the deep potential well
(formation of CO$_2$) on the underlying PES (cf., Figure 1 of Ref.
\citenum{VelizPCCP2021}) and the large masses of the participating
atoms make the quantum dynamical calculation computationally
challenging. The undulations in the total reaction probability (see
Figure \ref{fig:fig3}) are indicative of formation of intermediate
collision complexes inside the deep potential
well.\cite{jorfi2009state} However, the higher exoergicity of the
reaction facilitates breakup of these complexes and thus shortens
their lifetimes. Despite a deeper well on the underlying PES, lower
exoergicity could change these undulations into sharp and intense
resonance oscillations which was found for the S+OH, C+OH, C+H$_2$,
and S+H$_2$/D$_2$/HD
reactions.\cite{goswami2018theoretical,GoswamiMolPhys2017,goswami2014time,banares2003quantum,banares2005quantum}\\

\noindent
The reaction probabilities from the wavepacket simulations in Figure
\ref{fig:fig3} reveal that reactant vibrational excitation decreases
the reactivity at low and intermediate collision energies, whereas the
reactivity is enhanced for O$_{2}$($v=2$, $j=0$) at the highest
collision energies considered. Undulations for the total reaction
probability are also observed for vibrationally excited reactants and
analyzed further below.\\

\noindent
Next, the reaction probabilities from the QCT simulations are
considered and compared with those obtained from the TDWP
approach. QCT reaction probabilities are largest for O$_2 (v=0, j=0)$
and decay with increasing collision energy, see green (GB) and red
(HB) circles in the inset of Figure \ref{fig:fig3}. For O$_2 (v=1,
j=0)$ (blue circles) at low $E_{\rm c}$ the reaction probability is
smaller than for O$_2 (v=0, j=0)$ but approaches it with increasing
collision energy. Interestingly, for O$_2 (v=2, j=0)$ (brown circles)
the QCT reaction probability starts low at low $E_{\rm c}$, increases
- similar to TDWP - and then also decays towards a comparable
amplitude as for $v=1$ and $v=0$. Compared with the results from TDWP
simulations the QCT simulations are in excellent overall agreement
except for the lowest collision energies considered. In contrast to
TDWP, the QCT results decay monotonically for higher collision
energies also for O$_2 (v=0, j=0)$, (see inset of Figure
\ref{fig:fig3}). Because the wavepacket for the TDWP simulations was
optimized for O$_2 (v=0)$ and not for O$_2 (v=1)$ or O$_2 (v=2)$ it is
possible that the TDWP results are not fully converged at the higher
collision energies considered.\\

\subsection{Product Vibrational State Resolved Reaction Probabilities}
\begin{figure}
 \begin{center}
  \includegraphics*[scale=0.7]{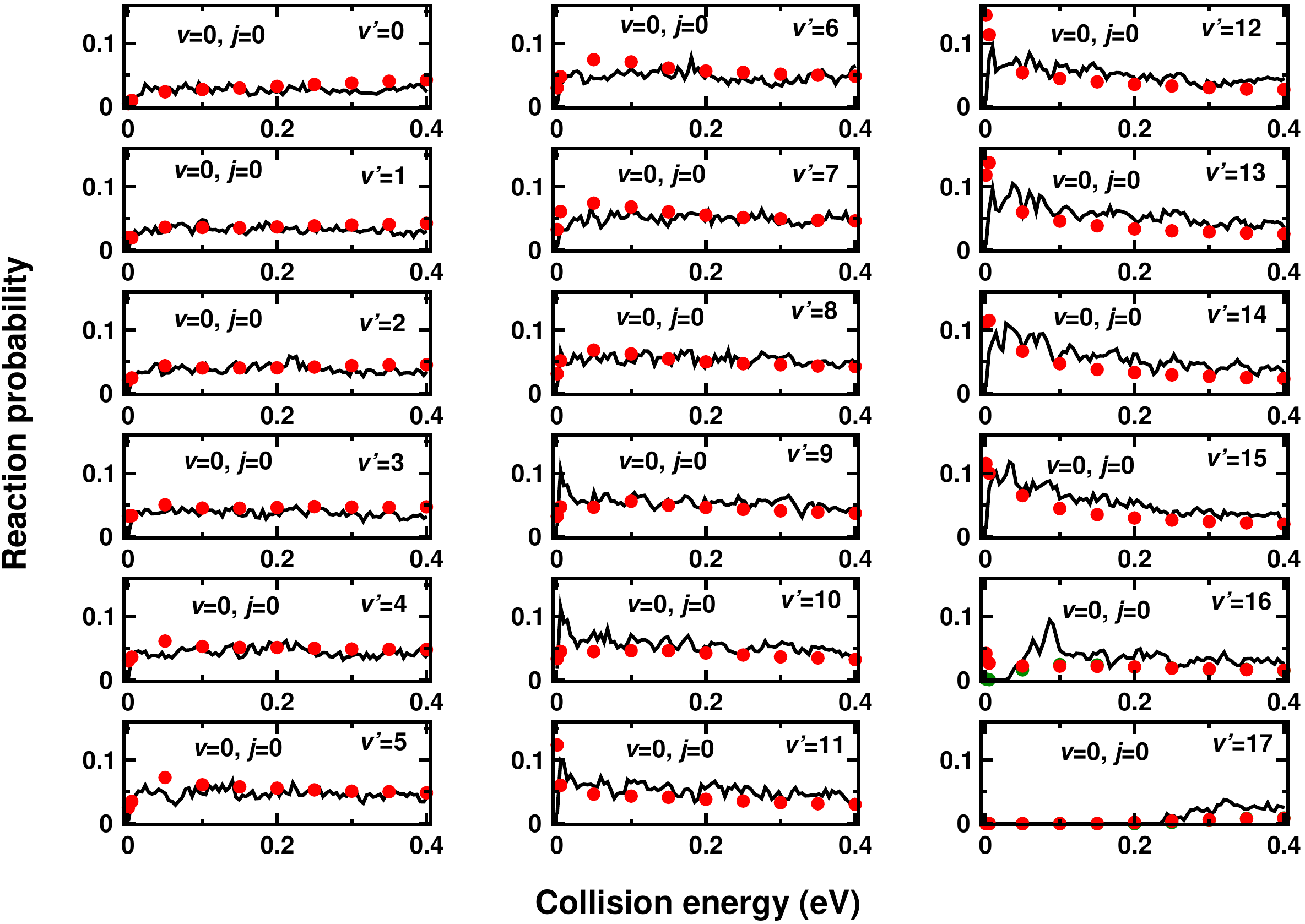}
  \caption{ Product diatom vibrational
    level resolved state-to-state reaction probabilities for the
    C($^{3}$P) + O$_{2}$($^3\Sigma_{g}^{-}$, $v=0$, $j=0$)
    $\rightarrow$ CO($^{1}\Sigma^{+}$, $v{'}$, $\sum j{'}$)+
    O($^{1}$D) reaction on its electronic ground state and for $J=0$
    as function of collision energy. The probabilities calculated from
    the TDWP approach and QCT(HB) method are shown by solid (black)
    lines and solid (red) dots, respectively. Probabilities are also
    calculated by using QCT(GB) method for $v{'}=16-17$ and the
    results are shown by solid green dots.}
\label{fig:fig4}
\end{center}
\end{figure}

Product vibrational level resolved reaction probabilities for
CO($v{'}=0-17$) and for $J=0$ are shown in Figure \ref{fig:fig4} as a
function of collision energy. The TDWP probabilities are the solid
(black) lines and the QCT(HB) probabilities are the red circles. For
comparison, probabilities are also determined from GB for $v{'}=16-17$
(solid green dots) in Figure \ref{fig:fig4}. Similar to the total
reaction probabilities, undulations are also found in product
vibrational level resolved TDWP probabilities, see Figures
\ref{fig:fig3} and \ref{fig:fig4}. The sharp increase of probability
at the low collision energies is not found for low $v{'}$. Conversely,
the TDWP probabilities for $v{'}=9-13$ possess significant qualitative
similarity with the total reaction probability.\\

\noindent
A comparison of the state-to-state TDWP and QCT reaction probabilities
in Figure \ref{fig:fig4} shows that the results are in excellent
agreement with each other at all collision energies considered for
$v{'}=0-10$, which is remarkable. However, differences at low
collision energies ($E_{\rm col} < 0.05$ eV) appear for
$v{'}=11-15$. Accumulation of these differences lead to the
differences for the total reaction probabilities from QCT and TDWP
simulations.\\

\noindent
The two panels of Figure \ref{fig:fig4} for CO($v{'}= 16, 17$) reveal
that TDWP find energy thresholds for product formation for $E_{\rm
  col} \sim 0.014$ and $\sim 0.22$ eV, respectively. This is in quite
good agreement with a value of $0.19$ eV for $v{'}=17$ from recent QCT
simulations.\cite{VelizPCCP2021} With histogram binning for analyzing
the QCT results there is no threshold for CO($v{'}= 16$) but
generating CO$(v'=17)$ requires collision energy. And analyzing the
same data with GB finds thresholds for CO($v{'}=16, 17$), see Figure
\ref{fig:fig4}. For the reaction with vibrational energy in the
reactant state (C + O$_{2}$($v=1$, $j=0$), see Figure \ref{sifig:fig1}
the product diatom vibrational level-resolved reaction probabilities
from TDWP and QCT simulations compare as well as for O$_{2}$($v=0$,
$j=0$). It can also be seen that reactant vibrational excitation has
an insignificant effect on product vibrational level resolved reaction
probabilities.\\

\subsection{Product Vibrational State Distributions}
\begin{figure}
 \begin{center}
  \includegraphics*[scale=0.7]{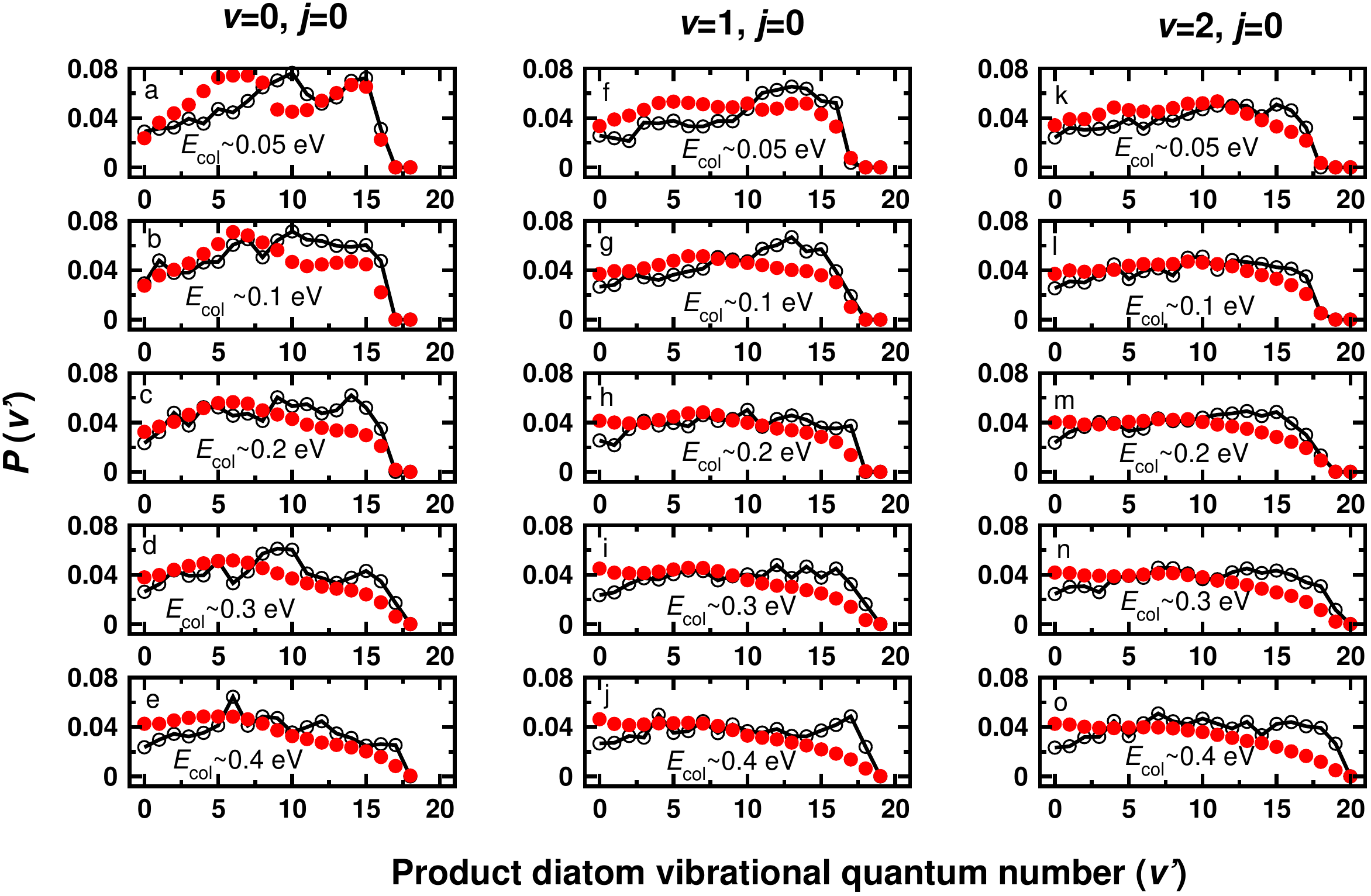}
  \caption{Product diatom vibrational level distributions at five
    different collision energies ($E_{\rm col}$) for the C +
    O$_{2}(v=0-2, j=0)$ reaction, by means of reaction probabilities
    of $J=$0. The results calculated by employing the TDWP and QCT
    approaches are shown by black lines with open circle and solid red
    dots, respectively. These QCT probabilities are obtained by using
    the HB method. The parameters of the TDWP approach were optimized
    for reactant O$_{2}(v=0, j= 0)$ and also used for the wave
    packets corresponding to vibrationally excited O$_{2}$.}
    \label{fig:fig8}
 \end{center}
\end{figure}

The product (CO) diatom vibrational level distributions $P(v')$ at
five different collision energies for the C + O$_{2}$($v=0-2$, $j=0$)
reaction with $J=0$ are shown in Figure \ref{fig:fig8} from TDWP (open
circles) and QCT (red dots) simulations. Overall, the final state
distributions from the two approaches agree favourably although
$P(v')$ from QCT simulations are smoother than those from the TDWP
simulations. This is particularly prevalent for low collision energies
and for initial $v=0$. With increasing vibrational excitation in the
reactant $P(v')$ from the TDWP simulations also become smoother and
the results from QCT overlap closely with them. Again, it may be that
the results from TDWP for initial $v=1$ and $v=2$ are slightly
affected by the fact that the initial wavepacket was optimized for
$v=0$. Compared with earlier work on the C+OH reaction the present
results indicate somewhat closer agreement between the TDWP and QCT
approaches, in particular with regards to pronounced minima and maxima
at intermediate values of $v'$ in $P(v')$ from the TDWP
simulations.\cite{bulut2009time} It is also noteworthy to mention here
that although the TDWP and QCT total reaction probabilities for
O$_{2}$($v=2$, $j=0$) differ at higher energies (Figure
\ref{fig:fig3}), the vibrational distributions at $E_{\rm col} \sim
0.3$ eV and 0.4 eV agree reasonably well except for high $v{'} \geq
15$, see Figure \ref{fig:fig8}.\\

\noindent

\noindent
It can also be seen from Figure \ref{fig:fig8} that with increasing
vibrational excitation of the reactant diatom the number of product
vibrational state increases. The energy supplied to reactant vibration
enhances the total energy of the system and makes the opening of
vibrationally excited product channels possible. This suggests that
part of the reactant internal energy is transferred/disposed into
product vibration. However, a careful look at Figure \ref{fig:fig8}
reveals that the collision energy and reactant vibrational excitation
hardly have any effect on the overall pattern of the distribution and
thus it becomes very difficult to find a general/regular trend for the
variation of the distribution with these two types of energy.\\

\subsection{Product rotational level distributions}
\begin{figure}
 \begin{center}
  \includegraphics*[scale=0.6]{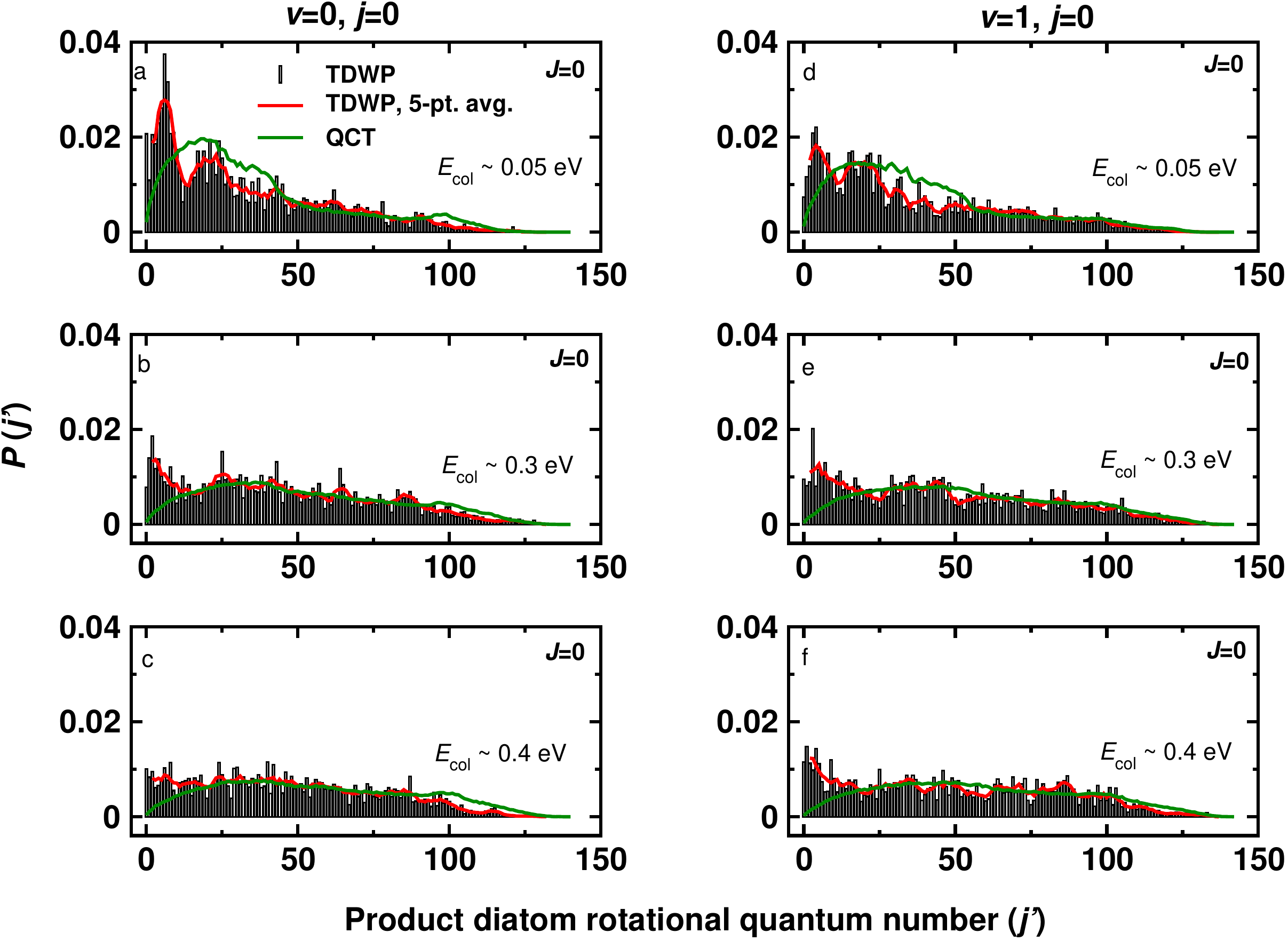}
  \caption{Product diatom rotational
    level distributions of the C + O$_{2}$($v=0-1$, $j=0$)
    $\rightarrow$ CO($\sum v{'}$, $j{'}$) + O reaction at three
    different collision energies ($E_{\rm col}$) as function of product
    diatom rotational quantum number $j{'}$. The TDWP and QCT results
    are shown in black bars and green line types, respectively. The
    red line in each panel represents a 5-point average of the TDWP
    distributions. The distributions for O$_{2}$($v=0$, $j=0$) and
    O$_{2}$($v=1$, $j=0$) are shown in panels a-c and d-f,
    respectively.}
    \label{fig:fig9}
 \end{center}
\end{figure}

The product diatom rotational level distributions $P(j')$ of the C +
O$_{2}$($v=0-1$, $j=0$) $\rightarrow$ CO($\sum v{'}$, $j{'}$) +O
reaction at three different collision energies ($E_{\rm col} \sim0.05$
eV, $0.3$ eV and $0.4$ eV) are shown from TDWP (black bars) and QCT
simulations (green lines) in Figures \ref{fig:fig9}a-f.  A 5-point
average (red) for the TDWP distribution in each panel allows for
easier comparison with the QCT results. Figure \ref{fig:fig9} shows
that the agreement of the TDWP and QCT rotational distributions is
excellent for moderate and high $j{'}$ and also improves with
increasing collision energy. Disagreement between the results from the
two approaches is found for low $j{'}$. The TDWP results predict a
peak at low $j{'}$ which is missing in the QCT results. A possible
explanation for this is the fact that low $j'$ from QCT simulation is
typically associated with separation of the two fragments from a
near-collinear geometry which is unlikely to occur given the
anisotropy of the PES. On the other hand, due to its spatial extent,
the wavepacket samples such regions. Given the small rotational
constant of the product, even small rotational energies can lead to
appreciable population of low-$j'$ states.\\

\noindent
It can also be seen from each row of Figure \ref{fig:fig9} that
reactant vibrational excitation does not change the pattern of the
distributions. This is observed in both TDWP and QCT results. Thus,
similar to the vibrational distributions (Figure \ref{fig:fig8}), the
overall pattern of the rotational distributions is also unaffected by
reactant vibrational excitation. However, the peak height of the TDWP
distributions at the low $j{'}$ gets lowered and the distribution
becomes flatter at $E_{\rm col} \sim 0.4$ eV. This is more prominent
when the reactant is in its ground ro-vibrational level. Therefore, at
higher collision energies the lower and moderate $j{'}$ levels of the
product diatom is formed with almost equal possibility.\\

\section{Discussion and Conclusion} 
The dynamics of the C + O$_{2}$ ($^{3}\Sigma_{g}^{-}$, $v= 0-2$,
$j=0$) $\rightarrow$ CO($^{1}\Sigma^{+}$, $v{'}$, $j{'}$) + O($^{1}$D)
reaction is investigated from TDWP and QCT calculations for $J=0$
considering the $^{1}$A${'}$ electronic ground state. The energy
resolved total reaction probabilities calculated by these two
approaches are in overall excellent agreement with each other except
for the lowest collision energies and for reactant O$_{2}$($v=2$,
$j=0$) at higher energies. Enhancement of the reactivity for
O$_{2}$($v=2$, $j=0$) at higher energies is obtained from the WP
calculations whereas reactivity predicted by QCT remains insensitive
to reactant vibrational excitation.  This qualitative disagreement can
be attributed to the possible non-convergence of the WP probabilities
at the higher energies for O$_{2}$($v=2$, $j=0$) because the
wavepacket was optimized for the O$_2(v=0)$ ground state.\\

\begin{figure}
 \begin{center}
  \includegraphics*[scale=0.6]{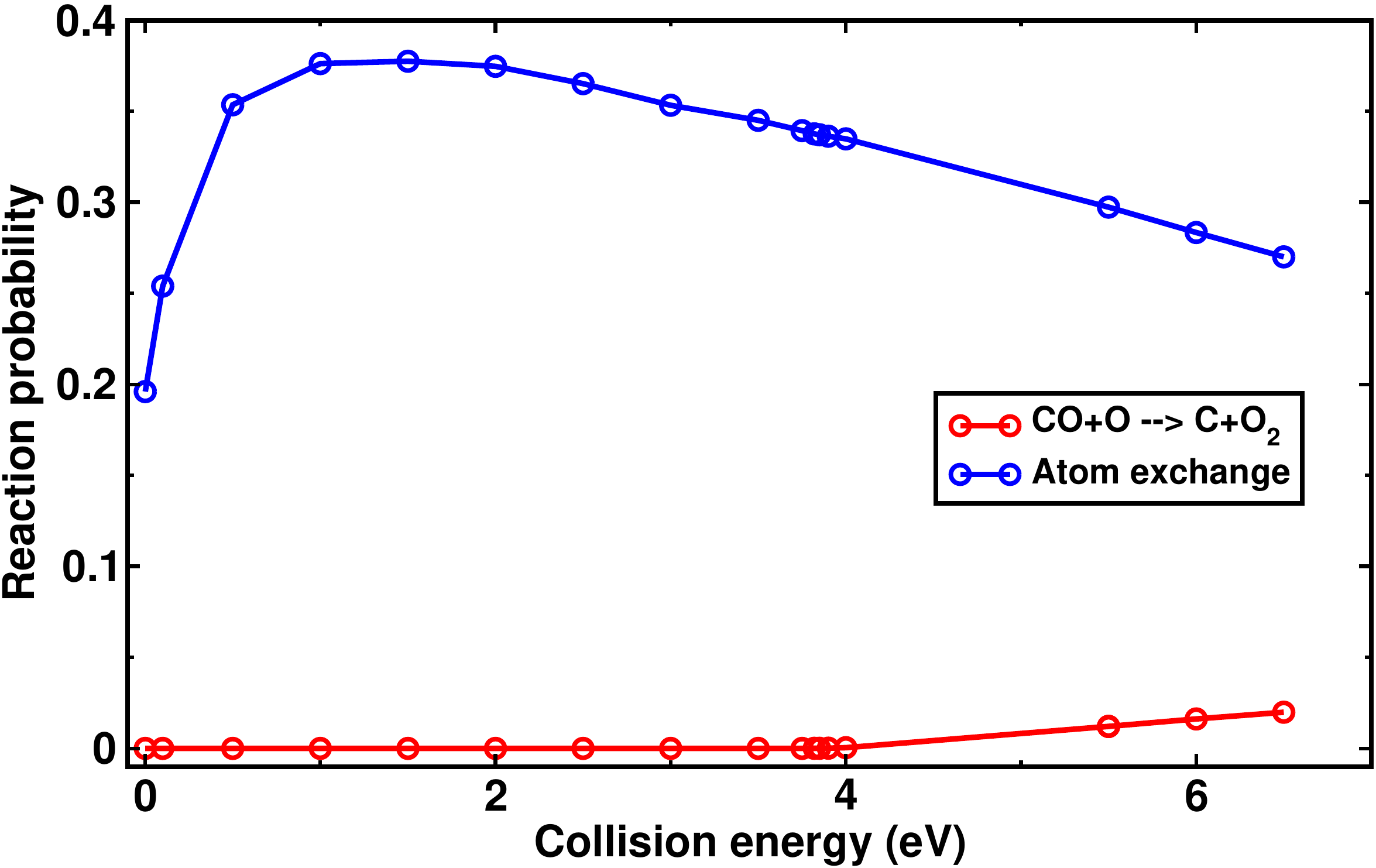}
  \caption{Energy resolved total reaction probabilities of the,
    CO$_{\rm A}$($v=0$, $j=0$) + O$_{\rm B}$ $\rightarrow$ C + O$_{\rm
      A}$O$_{\rm B}$ and CO$_{\rm B}$ + O$_{\rm A}$, reaction as
    function of collision energy and for $J$=0, calculated by
    employing the QCT approach.}
    \label{fig:fig10}
 \end{center}
\end{figure}

\noindent
In addition to the forward C + O$_{2}$ reaction, which was the topic
up to this point, the thermoneutral atom-exchange, CO$_{\rm A}$ +
O$_{\rm B}$ $\rightarrow$ CO$_{\rm B}$ + O$_{\rm A}$, and the
endoergic backward CO$_{\rm A}$ + O$_{\rm B}$ $\rightarrow$ C +
O$_{\rm A}$O$_{\rm B}$ reactions also take place on the $^{1}$A${'}$
PES. To complete the picture, total reaction probabilities from QCT
simulations for these two reactions were calculated up to $6.5$ eV
collision energy and the results are shown in Figure \ref{fig:fig10}
as function of collision energy. It can be seen from Figure
\ref{fig:fig10} that the CO$_{\rm A}$($v=0$, $j=0$) + O$_{\rm B}$
$\rightarrow$ C + O$_{\rm A}$O$_{\rm B}$ reaction has a high energy
threshold whereas the atom-exchange reaction is
barrierless. Comparison of probabilities shown in Figure
\ref{fig:fig3} and \ref{fig:fig10} reveals that the reactivity is
highest for the exoergic C + O$_{2} \rightarrow$ CO + O reaction,
followed by the atom-exchange reaction. The TDWP approach can also be
applied here, however, the parameters shown in Table \ref{tab:tab1}
need to be converged separately for the atom-exchange and endoergic
backward reactions which is outside the scope of the present
work. Given the good agreement between TDWP and QCT simulations it is,
however, expected that the results from Figure \ref{fig:fig10} are
representative.\\

\noindent
The thermoneutral atom-exchange reaction (CO$_{\rm A}$ + O$_{\rm B}$
$\rightarrow$ CO$_{\rm B}$ + O$_{\rm A}$) has already been considered
in the context of stabilizing CO$_2$ on amorphous solid water (ASW) to
better understand molecule formation at astrophysical
conditions.\cite{MM.co2:2021,MM.co2.2:2021} For this process it was
found that for bulk temperatures of $\sim 50$ K formation and
stabilization of CO$_2$ is very likely (60 \% of the trajectories
stabilize). Together with the present results this suggests that
quantum effects will only marginally affect this conclusion. This is
also consistent with results from QCT and TDWP simulations for
O($^3$P)+O($^3$P) recombination for which quantum mechanical effects
were also found to be far less important than effects of ASW surface
roughness on the diffusivity and recombination
dynamics.\cite{MM.o2:2018,MM.o2:2020}\\

\noindent
The undulations in the total reaction probabilities from TDWP
simulations (Figures \ref{fig:fig3}, \ref{fig:fig4} and
\ref{sifig:fig1}) are attributed to formation and transient
stabilization of the CO$_2$ intermediate inside the potential
well. These are largely unaffected by reactant vibrational excitation,
and merit some further discussion. Based on earlier
work\cite{jorfi2009state} these undulations can be associated with
formation of a transiently stabilized CO$_2$ intermediate. Large
undulations with some resonances were found in the QM total reaction
probabilities of the exoergic and barrierless N + OH $\rightarrow$ NO
+ H reaction proceeding through deep HON well on the underlying
PES.\cite{jorfi2009state} Furthermore, dense resonance structures are
found between $0.41$ and $0.43$ eV collision
energies.\cite{jorfi2009state} The authors have associated these
resonances to the formation of long-lived intermediate complex inside
the HON potential well.\cite{jorfi2009state}. The exoergicities of the
N + OH and present C + O$_{2}$ reactions are $1.99$ eV and $\sim 3.8$
eV. Therefore, large undulations are found in the reaction
probabilities of the present C + O$_{2}$ reaction due to the formation
of transiently stabilized CO$_{2}$ intermediate inside the potential
well on the underlying PES.\\

\noindent
The black-solid curve in Figure \ref{fig:fig3} shows that there are
$19$ prominent peaks across the entire range of collision energies
considered. Peak separations range from $\sim 0.009$ eV to $\sim
0.036$ eV with an average of $\sim 0.021$ eV. Using the energy/time
uncertainty relationship $\Delta E \times \Delta t = \hbar$,
corresponding lifetimes range from $\sim 0.018$ ps to $\sim 0.073$ ps
with an average of $\sim 0.038$ ps.\\

\begin{figure}
  \begin{center}
   \includegraphics*[scale=0.5]{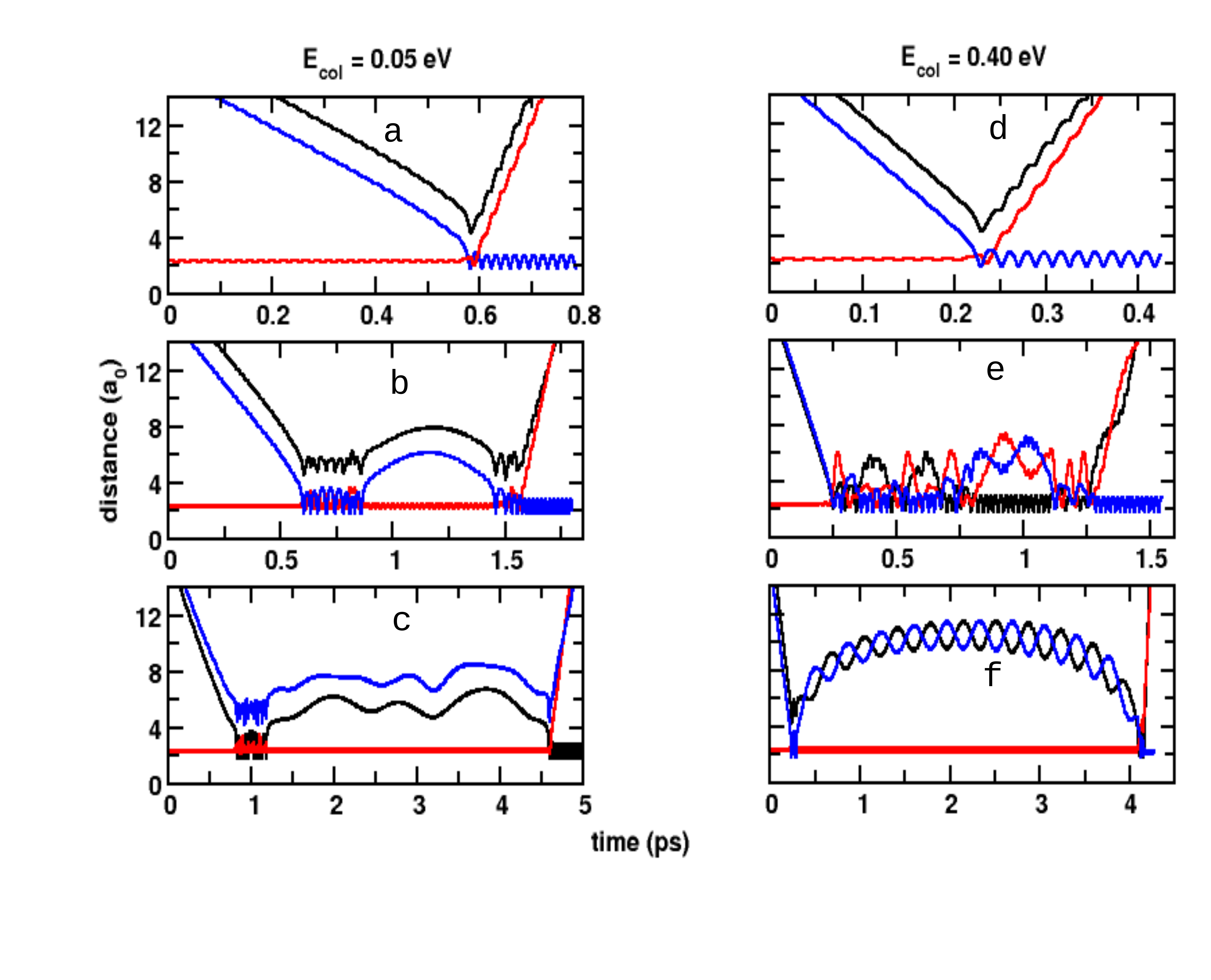}
   \caption{Time evolution of the inter nuclear distances of the three
     diatoms, O$_{2}$ (red), CO$_{\rm A}$ (blue) and CO$_{\rm B}$
     (black) as function of time for $E_{\rm col}=$ 0.05 eV (left
     column) and $E_{\rm col}=$ 0.40 eV (right column). Collision
     times are $\sim 0.03$, $\sim 1.0$, $\sim3.8$, $\sim 0.04$,
     $\sim1.02$ and $\sim3.9$ ps.}
    \label{fig:fig7}
\end{center}
\end{figure}

\begin{figure}
 \begin{center}
  \includegraphics*[scale=0.6]{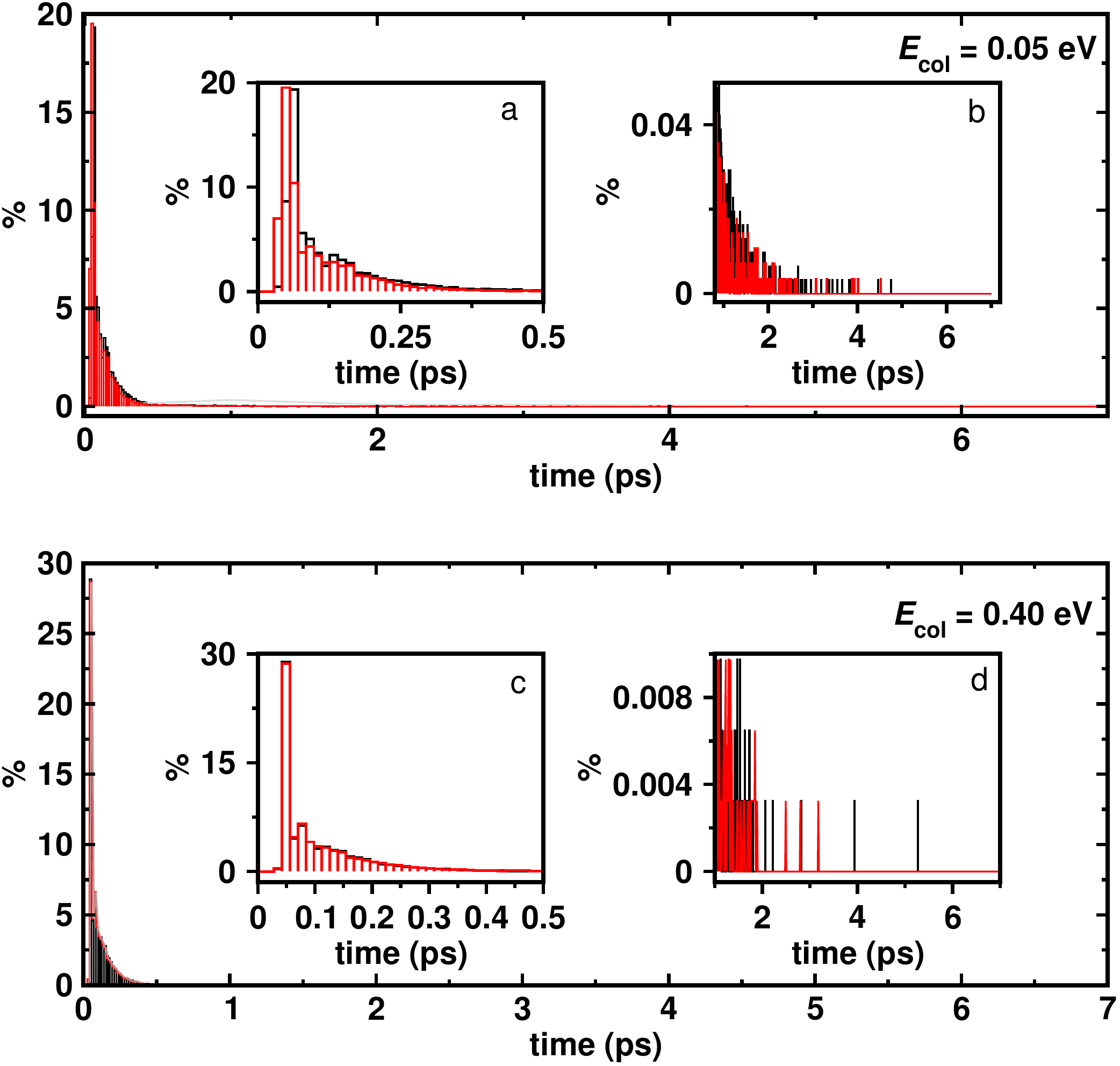}
  \caption{Distributions of collision time (or lifetime) of the
    intermediate collision complex as function of time analyzed with a
    cutoff of 14 a$_0$ (black) and 12 a$_0$ (red) for the sum of
    distances as the criterion for formation of the complex. The
    distributions at $0.05$ eV and $0.40$ eV collision energies are
    shown in the top and bottom panels, respectively, together with
    detailed views for short and long collision times in the insets.
    The maxima for the two distributions with a distance cutoff at 14
    a$_0$ are at $\sim 0.07$ ps and $\sim 0.056$ ps and the longest
    collision times encountered are $\sim 4.8$ ps and $\sim 5.3$ ps
    for the two cases considered.  These values change to $\sim 0.056$
    ps and $\sim 0.056$ ps and $\sim 4.5$ ps and $\sim 3.2$ ps for the
    maxima and longest collision time, respectively, if a cutoff of 12
    a$_0$ is used.}
    \label{fig:fig6}
\end{center}
\end{figure}

\noindent
The three internuclear separations from representative QCT simulations
for $E_{\rm col} = 0.05$ eV and $E_{\rm col} = 0.40$ eV are shown in
Figure \ref{fig:fig7} and the corresponding collision time
distributions $P(\tau_{\rm c})$ are reported in Figure
\ref{fig:fig6}. Here, $\tau_{\rm c}$ is defined as the time for which
the sum of the three internuclear distances is smaller than 14 a$_0$
or 12 a$_0$, respectively. The distribution $P(\tau_{\rm c})$ was
determined from 22000 reactive trajectories at each of the collision
energies. The time series in Figures \ref{fig:fig7}a and d represent
trajectories with short collision time $\tau_{\rm c} \sim 0.03$ ps and
$\sim0.04$ ps, respectively, whereas Figures \ref{fig:fig7}b and e
show trajectories with longer $\tau_{\rm c}$ ($\sim1.0$ ps and
$\sim1.02$ ps, respectively). The results of panels b and e clearly
indicate the formation of triatomic intermediate complexes whereas the
findings of panels c and f suggest the formation of weakly bound
triatomic complexes with short contact times, reminiscent of
roaming.\\

\noindent
The collision time distributions from QCT simulations are strongly
peaked with their maxima at $\tau_{\rm c} \sim 0.07$ ps and $\sim
0.056$ ps for $E_{\rm col} = 0.05$ eV and 0.40 eV, respectively, see
inset of Figure \ref{fig:fig6}. The probability for finding such short
collision times is 20 \% and 30 \%, respectively, for the two
collision energies considered. If a shorter distance cutoff of 12
a$_0$ is used the maxima of $P(\tau_c)$ are at $\tau_{\rm c} = 0.056$
ps whereas the longest lifetimes are 4.5 ps and 3.2 ps for the two
collision energies. Hence, the analysis of the lifetimes is only
moderately affected by the choice of the cutoff value for the sum of
the three interatomic distances.\\

\noindent
Collision times of the order of 0.06 ps are in quite good qualitative
agreement with the analysis of the recurrences in the reaction
probabilities from TDWP simulations which yielded lifetimes of around
$\sim 0.04$ ps, see above. On the other hand, the QCT simulations also
find a long tail in the collision time distributions extending out to
$\sim 4.8$ ps and $\sim 5.3$ ps for the longest collision times for
the two simulation conditions considered. Overall, it is found that
the implied time scales of the undulations in the TDWP reaction
probabilities, which have previously been linked to formation of a
collisionally stabilized intermediate,\cite{jorfi2009state} are
consistent with the collision times from the QCT simulations, and that
they constitute an inherent characteristic of the reaction
considered.\\

\noindent
The agreement between the TDWP and QCT product vibrational level
resolved probabilities for O$_{2}$($v=0-1$, $j=0$) is remarkably good
except for low $E_{\rm col}$ and only for specific values of
$v{'}$. Reactant vibrational excitation does not change the pattern of
these state-to-state probabilities significantly. For the product
rotational level distributions at different collision energies, the
agreement between TDWP and QCT results is excellent for moderate and
higher $j' \gtrsim 15$ values and improves with increasing collision
energy. The peak in $P(j')$ from TDWP simulations at low $j{'}$ is not
obtained from the QCT calculations and reactant vibrational excitation
has little impact on the overall behaviour of the final state
distribution.\\

\noindent
The present results can also be compared with results from pulsed,
crossed, supersonic molecular beams, at relative translational
energies between 4.4 and 90 meV.\cite{costes:1998} Such experiments
found that collision energy $E_{\rm c} > 0.04$ eV is required to open
the product CO($v' = 17$) channel. For CO($v{'}=17$), the present QCT
investigation shows a threshold of $\sim 0.15$ eV from analysis based
on histogram and $\sim 0.25$ eV from Gaussian binning,
respectively. An earlier QCT investigation\cite{VelizPCCP2021} on a
finer grid for $E_{\rm c}$ showed that for ($v=0$, $j<10$) the
threshold is $0.18$ eV which changes to 0.25 eV if the same data is
analyzed for ($v=0$, $j=0$). This is also consistent with $\sim 0.22$
eV from the TDWP simulations. Including higher rotational states for
the reactant progressively shift the threshold to 0.04 eV when all
reactant $j-$states are included which agrees well with the
experiments.\cite{costes:1998,VelizPCCP2021}\\

\noindent
In summary, this work provides a stringent comparison for total and
product diatom vibrational state resolved reaction probabilities and
product state vibrational and rotational distributions from TDWP and
QCT simulations for the C+O$_2$ $\rightarrow$ O+CO reaction on its
ground electronic state. Overall, the agreement of a quasi-classical
description with results from wavepacket simulations is encouraging
except for the lowest collision energies or lowest product rotational
states. Expected quantum effects such as undulations in the total
reaction probability $P(E_c)$ can be rationalized in a qualitative
fashion from analysis of the QCT trajectories. Given the ease with
which QCT simulations can be carried out the present work suggests
that they are a meaningful approach for understanding the dynamics of
atom plus diatom reactions and for larger systems as well.\\

\section*{Acknowledgments}
Support by the Swiss National Science Foundation through grants
200021-117810, the NCCR MUST (to MM), and the University of Basel is
acknowledged. Part of this work was supported by the United State
Department of the Air Force, which is gratefully acknowledged (to
MM). This project has received funding from the European Union Horizon
2020 funding scheme under MCSA Grant No 801459, FP-RESOMUS.

\bibliography{refs.tidy}

\providecommand{\latin}[1]{#1}
\makeatletter
\providecommand{\doi}
  {\begingroup\let\do\@makeother\dospecials
  \catcode`\{=1 \catcode`\}=2 \doi@aux}
\providecommand{\doi@aux}[1]{\endgroup\texttt{#1}}
\makeatother
\providecommand*\mcitethebibliography{\thebibliography}
\csname @ifundefined\endcsname{endmcitethebibliography}
  {\let\endmcitethebibliography\endthebibliography}{}
\begin{mcitethebibliography}{75}
\providecommand*\natexlab[1]{#1}
\providecommand*\mciteSetBstSublistMode[1]{}
\providecommand*\mciteSetBstMaxWidthForm[2]{}
\providecommand*\mciteBstWouldAddEndPuncttrue
  {\def\EndOfBibitem{\unskip.}}
\providecommand*\mciteBstWouldAddEndPunctfalse
  {\let\EndOfBibitem\relax}
\providecommand*\mciteSetBstMidEndSepPunct[3]{}
\providecommand*\mciteSetBstSublistLabelBeginEnd[3]{}
\providecommand*\EndOfBibitem{}
\mciteSetBstSublistMode{f}
\mciteSetBstMaxWidthForm{subitem}{(\alph{mcitesubitemcount})}
\mciteSetBstSublistLabelBeginEnd
  {\mcitemaxwidthsubitemform\space}
  {\relax}
  {\relax}

\bibitem[Sharma \latin{et~al.}(2010)Sharma, Swantek, Flaherty, Austin,
  Doraiswamy, and Candler]{Sharma2010}
Sharma,~M.; Swantek,~A.~B.; Flaherty,~W.; Austin,~J.~M.; Doraiswamy,~S.;
  Candler,~G.~V. Experimental and Numerical Investigation of Hypervelocity
  Carbon Dioxide Flow over Blunt Bodies. \emph{J. Thermophys. Heat Transf.}
  \textbf{2010}, \emph{24}, 673--683\relax
\mciteBstWouldAddEndPuncttrue
\mciteSetBstMidEndSepPunct{\mcitedefaultmidpunct}
{\mcitedefaultendpunct}{\mcitedefaultseppunct}\relax
\EndOfBibitem
\bibitem[Husain and Young(1975)Husain, and Young]{husain:1975}
Husain,~D.; Young,~A.~N. Kinetic investigation of ground state carbon atoms{,}
  C(2$^3$P). \emph{J. Chem. Soc.{,} Faraday Trans. 2} \textbf{1975}, \emph{71},
  525--531\relax
\mciteBstWouldAddEndPuncttrue
\mciteSetBstMidEndSepPunct{\mcitedefaultmidpunct}
{\mcitedefaultendpunct}{\mcitedefaultseppunct}\relax
\EndOfBibitem
\bibitem[Bergeat \latin{et~al.}(1999)Bergeat, Calvo, Dorthe, and
  Loison]{BERGEAT1999}
Bergeat,~A.; Calvo,~T.; Dorthe,~G.; Loison,~J. Fast-flow study of the C+NO and
  C+O$_2$ reactions. \emph{Chem.\ Phys.\ Lett.} \textbf{1999}, \emph{308},
  7--12\relax
\mciteBstWouldAddEndPuncttrue
\mciteSetBstMidEndSepPunct{\mcitedefaultmidpunct}
{\mcitedefaultendpunct}{\mcitedefaultseppunct}\relax
\EndOfBibitem
\bibitem[Geppert \latin{et~al.}(2000)Geppert, Reignier, Stoecklin, Naulin,
  Costes, Chastaing, Le~Picard, Sims, and Smith]{Geppert2000}
Geppert,~W.~D.; Reignier,~D.; Stoecklin,~T.; Naulin,~C.; Costes,~M.;
  Chastaing,~D.; Le~Picard,~S.~D.; Sims,~I.~R.; Smith,~I. W.~M. Comparison of
  the cross-sections and thermal rate constants for the reactions of C($^3$P)
  atoms with O$_2$ and NO. \emph{Phys. Chem. Chem. Phys.} \textbf{2000},
  \emph{2}, 2873--2881\relax
\mciteBstWouldAddEndPuncttrue
\mciteSetBstMidEndSepPunct{\mcitedefaultmidpunct}
{\mcitedefaultendpunct}{\mcitedefaultseppunct}\relax
\EndOfBibitem
\bibitem[Becker \latin{et~al.}(1988)Becker, Brockmann, and Weisen]{becker:1988}
Becker,~K.~H.; Brockmann,~K.~J.; Weisen,~P. Spectroscopic identification of
  C($^3$P) atoms in halogenomethane + H flame systems and measurements of
  C($^3$P) reaction rate constants by two-photon laser-induced fluorescence.
  \emph{J. Chem. Soc.{} Faraday Trans. 2} \textbf{1988}, \emph{84},
  455--461\relax
\mciteBstWouldAddEndPuncttrue
\mciteSetBstMidEndSepPunct{\mcitedefaultmidpunct}
{\mcitedefaultendpunct}{\mcitedefaultseppunct}\relax
\EndOfBibitem
\bibitem[Dorthe \latin{et~al.}(1991)Dorthe, Caubet, Vias, Barrere, and
  Marchais]{dorthe:1991}
Dorthe,~G.; Caubet,~P.; Vias,~T.; Barrere,~B.; Marchais,~J. Fast flow studies
  of atomic carbon kinetics at room temperature. \emph{J.~Phys.\ Chem.}
  \textbf{1991}, \emph{95}, 5109--5116\relax
\mciteBstWouldAddEndPuncttrue
\mciteSetBstMidEndSepPunct{\mcitedefaultmidpunct}
{\mcitedefaultendpunct}{\mcitedefaultseppunct}\relax
\EndOfBibitem
\bibitem[Chastaing \latin{et~al.}(1999)Chastaing, L.~James, R.~Sims, and
  W.~M.~Smith]{chastaing:1999}
Chastaing,~D.; L.~James,~P.; R.~Sims,~I.; W.~M.~Smith,~I. Neutral–neutral
  reactions at the temperatures of interstellar clouds: Rate coefficients for
  reactions of atomic carbon{,} C($^3$P){,} with O$_2${,} C$_2$H$_2${,}
  C$_2$H$_4$ and C$_3$H$_6$ down to 15 K. \emph{Phys. Chem. Chem. Phys.}
  \textbf{1999}, \emph{1}, 2247--2256\relax
\mciteBstWouldAddEndPuncttrue
\mciteSetBstMidEndSepPunct{\mcitedefaultmidpunct}
{\mcitedefaultendpunct}{\mcitedefaultseppunct}\relax
\EndOfBibitem
\bibitem[Chastaing \latin{et~al.}(2000)Chastaing, Le~Picard, and
  Sims]{chastaing:2000}
Chastaing,~D.; Le~Picard,~S.~D.; Sims,~I.~R. Direct kinetic measurements on
  reactions of atomic carbon, C($^3$P), with O$_2$ and NO at temperatures down
  to 15 K. \emph{J.~Chem.\ Phys.} \textbf{2000}, \emph{112}, 8466--8469\relax
\mciteBstWouldAddEndPuncttrue
\mciteSetBstMidEndSepPunct{\mcitedefaultmidpunct}
{\mcitedefaultendpunct}{\mcitedefaultseppunct}\relax
\EndOfBibitem
\bibitem[Dean \latin{et~al.}(1991)Dean, Davidson, and Hanson]{Hanson:1991}
Dean,~A.~J.; Davidson,~D.~F.; Hanson,~R.~K. A shock tube study of reactions of
  C atoms with H$_{2}$ and O$_{2}$ using excimer photolysis of C$_{3}$O$_{2}$
  and C atom atomic resonance absorption spectroscopy. \emph{J.~Phys.\ Chem.}
  \textbf{1991}, \emph{95}, 183--191\relax
\mciteBstWouldAddEndPuncttrue
\mciteSetBstMidEndSepPunct{\mcitedefaultmidpunct}
{\mcitedefaultendpunct}{\mcitedefaultseppunct}\relax
\EndOfBibitem
\bibitem[Ogryzlo \latin{et~al.}(1973)Ogryzlo, Reilly, and Thrush]{thrush:1973}
Ogryzlo,~E.; Reilly,~J.; Thrush,~B. Vibrational excitation of CO from the
  reaction C+ O$_2$. \emph{Chem.\ Phys.\ Lett.} \textbf{1973}, \emph{23},
  37--39\relax
\mciteBstWouldAddEndPuncttrue
\mciteSetBstMidEndSepPunct{\mcitedefaultmidpunct}
{\mcitedefaultendpunct}{\mcitedefaultseppunct}\relax
\EndOfBibitem
\bibitem[Dubrin \latin{et~al.}(1964)Dubrin, MacKay, Pandow, and
  Wolfgang]{dubrin:1964}
Dubrin,~J.; MacKay,~C.; Pandow,~M.; Wolfgang,~R. Reactions of atomic carbon
  with $\pi$-bonded inorganic molecules. \emph{J. Inorg. Nuc. Chem.}
  \textbf{1964}, \emph{26}, 2113--2122\relax
\mciteBstWouldAddEndPuncttrue
\mciteSetBstMidEndSepPunct{\mcitedefaultmidpunct}
{\mcitedefaultendpunct}{\mcitedefaultseppunct}\relax
\EndOfBibitem
\bibitem[Xantheas and Ruedenberg(1994)Xantheas, and Ruedenberg]{Xantheas1994}
Xantheas,~S.~S.; Ruedenberg,~K. Potential energy surfaces of carbon dioxide.
  \emph{Int. J. Quantum Chem.} \textbf{1994}, \emph{49}, 409--427\relax
\mciteBstWouldAddEndPuncttrue
\mciteSetBstMidEndSepPunct{\mcitedefaultmidpunct}
{\mcitedefaultendpunct}{\mcitedefaultseppunct}\relax
\EndOfBibitem
\bibitem[Jasper and Dawes(2013)Jasper, and Dawes]{Jasper2013}
Jasper,~A.~W.; Dawes,~R. Non-Born–Oppenheimer molecular dynamics of the
  spin-forbidden reaction O($^3$P) + CO(X$^1 \Sigma^+$) $\rightarrow$
  CO$_2$(X$^{1} \Sigma_{g}^{+}$). \emph{J.~Chem.\ Phys.} \textbf{2013},
  \emph{139}, 154313\relax
\mciteBstWouldAddEndPuncttrue
\mciteSetBstMidEndSepPunct{\mcitedefaultmidpunct}
{\mcitedefaultendpunct}{\mcitedefaultseppunct}\relax
\EndOfBibitem
\bibitem[Troe(1975)]{troe:1975}
Troe,~J. Thermal dissociation and recombination of polyatomic molecules.
  \emph{Fifth Symp. (Int.) Combust.} \textbf{1975}, \emph{15}, 667--679\relax
\mciteBstWouldAddEndPuncttrue
\mciteSetBstMidEndSepPunct{\mcitedefaultmidpunct}
{\mcitedefaultendpunct}{\mcitedefaultseppunct}\relax
\EndOfBibitem
\bibitem[Braunstein and Duff(2000)Braunstein, and Duff]{Braunstein2000}
Braunstein,~M.; Duff,~J.~W. Electronic structure and dynamics of O($^3$P) +
  CO($^1 \Sigma^+$) collisions. \emph{J.~Chem.\ Phys.} \textbf{2000},
  \emph{112}, 2736--2745\relax
\mciteBstWouldAddEndPuncttrue
\mciteSetBstMidEndSepPunct{\mcitedefaultmidpunct}
{\mcitedefaultendpunct}{\mcitedefaultseppunct}\relax
\EndOfBibitem
\bibitem[Kozlov \latin{et~al.}(2000)Kozlov, Makarov, Pavlov, and
  Shatalov]{shatalov:2000}
Kozlov,~P.; Makarov,~V.; Pavlov,~V.; Shatalov,~O. Experimental investigation of
  CO vibrational deactivation in a supersonic cooling gas flow. \emph{Shock
  Waves} \textbf{2000}, \emph{10}, 191--195\relax
\mciteBstWouldAddEndPuncttrue
\mciteSetBstMidEndSepPunct{\mcitedefaultmidpunct}
{\mcitedefaultendpunct}{\mcitedefaultseppunct}\relax
\EndOfBibitem
\bibitem[Center(1973)]{Center1973}
Center,~R.~E. Vibrational relaxation of {CO} by O atoms. \emph{J.~Chem.\ Phys.}
  \textbf{1973}, \emph{58}, 5230--5236\relax
\mciteBstWouldAddEndPuncttrue
\mciteSetBstMidEndSepPunct{\mcitedefaultmidpunct}
{\mcitedefaultendpunct}{\mcitedefaultseppunct}\relax
\EndOfBibitem
\bibitem[Kelley and Thommarson(1977)Kelley, and Thommarson]{Kelley1977}
Kelley,~J.~D.; Thommarson,~R.~L. Vibrational deactivation and atom exchange in
  O($^3$P) + CO(X$^1 \Sigma ^+$) collisions. \emph{J.~Chem.\ Phys.}
  \textbf{1977}, \emph{66}, 1953--1959\relax
\mciteBstWouldAddEndPuncttrue
\mciteSetBstMidEndSepPunct{\mcitedefaultmidpunct}
{\mcitedefaultendpunct}{\mcitedefaultseppunct}\relax
\EndOfBibitem
\bibitem[Davidson \latin{et~al.}(1978)Davidson, Schiff, Brown, and
  Howard]{Davidson1978}
Davidson,~J.~A.; Schiff,~H.~I.; Brown,~T.~J.; Howard,~C.~J. Temperature
  dependence of the deactivation of O($^{1}$D) by CO from 113-333 K.
  \emph{J.~Chem.\ Phys.} \textbf{1978}, \emph{69}, 1216--1217\relax
\mciteBstWouldAddEndPuncttrue
\mciteSetBstMidEndSepPunct{\mcitedefaultmidpunct}
{\mcitedefaultendpunct}{\mcitedefaultseppunct}\relax
\EndOfBibitem
\bibitem[Tully(1975)]{Tully:1975}
Tully,~J.~C. Reactions of O($^1$D) with atmospheric molecules. \emph{J.~Chem.\
  Phys.} \textbf{1975}, \emph{62}, 1893--1898\relax
\mciteBstWouldAddEndPuncttrue
\mciteSetBstMidEndSepPunct{\mcitedefaultmidpunct}
{\mcitedefaultendpunct}{\mcitedefaultseppunct}\relax
\EndOfBibitem
\bibitem[San Vicente~Veliz \latin{et~al.}(2021)San Vicente~Veliz, Koner,
  Schwilk, Bemish, and Meuwly]{VelizPCCP2021}
San Vicente~Veliz,~J.~C.; Koner,~D.; Schwilk,~M.; Bemish,~R.~J.; Meuwly,~M. The
  C($^{3}$P) + O$_{2}$($^{3}\Sigma_{g}^{-}$) $\longleftrightarrow$ CO$_{2}$
  $\longleftrightarrow$ CO($^{1}\Sigma^{+}$) + O($^{1}$D)/O($^{3}$P) reaction:
  thermal and vibrational relaxation rates from 15 K to 20000 K. \emph{Phys.
  Chem. Chem. Phys.} \textbf{2021}, \emph{23}, 11251--11263\relax
\mciteBstWouldAddEndPuncttrue
\mciteSetBstMidEndSepPunct{\mcitedefaultmidpunct}
{\mcitedefaultendpunct}{\mcitedefaultseppunct}\relax
\EndOfBibitem
\bibitem[Winter \latin{et~al.}(1973)Winter, Bender, and
  Goddard~III]{winter:1973}
Winter,~N.~W.; Bender,~C.~F.; Goddard~III,~W.~A. Theoretical assignments of the
  low-lying electronic states of carbon dioxide. \emph{Chem.\ Phys.\ Lett.}
  \textbf{1973}, \emph{20}, 489--492\relax
\mciteBstWouldAddEndPuncttrue
\mciteSetBstMidEndSepPunct{\mcitedefaultmidpunct}
{\mcitedefaultendpunct}{\mcitedefaultseppunct}\relax
\EndOfBibitem
\bibitem[Grebenshchikov and Borrelli(2012)Grebenshchikov, and
  Borrelli]{greben:2012}
Grebenshchikov,~S.~Y.; Borrelli,~R. Crossing Electronic States in the
  Franck-Condon Zone of Carbon Dioxide: A Five-Fold Closed Seam of Conical and
  Glancing Intersections. \emph{J.~Phys.\ Chem. Lett.} \textbf{2012}, \emph{3},
  3223--3227\relax
\mciteBstWouldAddEndPuncttrue
\mciteSetBstMidEndSepPunct{\mcitedefaultmidpunct}
{\mcitedefaultendpunct}{\mcitedefaultseppunct}\relax
\EndOfBibitem
\bibitem[Grebenshchikov(2013)]{greben:2013}
Grebenshchikov,~S.~Y. Photodissociation of carbon dioxide in singlet valence
  electronic states. II. Five state absorption spectrum and vibronic
  assignment. \emph{J.~Chem.\ Phys.} \textbf{2013}, \emph{138}, 224107\relax
\mciteBstWouldAddEndPuncttrue
\mciteSetBstMidEndSepPunct{\mcitedefaultmidpunct}
{\mcitedefaultendpunct}{\mcitedefaultseppunct}\relax
\EndOfBibitem
\bibitem[Grebenshchikov(2013)]{greben:2013.2}
Grebenshchikov,~S.~Y. Photodissociation of carbon dioxide in singlet valence
  electronic states. I. Six multiply intersecting ab initio potential energy
  surfaces. \emph{J.~Chem.\ Phys.} \textbf{2013}, \emph{138}, 224106\relax
\mciteBstWouldAddEndPuncttrue
\mciteSetBstMidEndSepPunct{\mcitedefaultmidpunct}
{\mcitedefaultendpunct}{\mcitedefaultseppunct}\relax
\EndOfBibitem
\bibitem[Schmidt \latin{et~al.}(2013)Schmidt, Johnson, and
  Schinke]{Schmidt2013}
Schmidt,~J.~A.; Johnson,~M.~S.; Schinke,~R. Carbon dioxide photolysis from 150
  to 210 nm: Singlet and triplet channel dynamics, UV-spectrum, and isotope
  effects. \emph{Proc. Natl. Acad. Sci.} \textbf{2013}, \emph{110},
  17691--17696\relax
\mciteBstWouldAddEndPuncttrue
\mciteSetBstMidEndSepPunct{\mcitedefaultmidpunct}
{\mcitedefaultendpunct}{\mcitedefaultseppunct}\relax
\EndOfBibitem
\bibitem[Zhou \latin{et~al.}(2013)Zhou, Zhu, Wen, Jiang, Yu, Lee, and
  Lin]{hsien:2013}
Zhou,~B.; Zhu,~C.; Wen,~Z.; Jiang,~Z.; Yu,~J.; Lee,~Y.-P.; Lin,~S.~H. Topology
  of conical/surface intersections among five low-lying electronic states of
  CO$_2$: Multireference configuration interaction calculations.
  \emph{J.~Chem.\ Phys.} \textbf{2013}, \emph{139}\relax
\mciteBstWouldAddEndPuncttrue
\mciteSetBstMidEndSepPunct{\mcitedefaultmidpunct}
{\mcitedefaultendpunct}{\mcitedefaultseppunct}\relax
\EndOfBibitem
\bibitem[Kinnersly and Murrell(1977)Kinnersly, and Murrell]{murrell:1977}
Kinnersly,~S.; Murrell,~J. A classical dynamical study of the reaction between
  C($^3$P) and O$_2$($^3 \Sigma_{g}^{-}$). \emph{Mol.\ Phys.} \textbf{1977},
  \emph{33}, 1479--1494\relax
\mciteBstWouldAddEndPuncttrue
\mciteSetBstMidEndSepPunct{\mcitedefaultmidpunct}
{\mcitedefaultendpunct}{\mcitedefaultseppunct}\relax
\EndOfBibitem
\bibitem[Brunsvold \latin{et~al.}(2008)Brunsvold, Upadhyaya, Zhang, Cooper,
  Minton, Braunstein, and Duff]{Brunsvold2008}
Brunsvold,~A.~L.; Upadhyaya,~H.~P.; Zhang,~J.; Cooper,~R.; Minton,~T.~K.;
  Braunstein,~M.; Duff,~J.~W. Dynamics of Hyperthermal Collisions of O($^3$P)
  with CO. \emph{J.~Phys.\ Chem.~A} \textbf{2008}, \emph{112}, 2192--2205\relax
\mciteBstWouldAddEndPuncttrue
\mciteSetBstMidEndSepPunct{\mcitedefaultmidpunct}
{\mcitedefaultendpunct}{\mcitedefaultseppunct}\relax
\EndOfBibitem
\bibitem[Koner \latin{et~al.}(2020)Koner, Bemish, and Meuwly]{MM.rev:2020}
Koner,~D.; Bemish,~R.~J.; Meuwly,~M. Dynamics on multiple potential energy
  surfaces: quantitative studies of elementary processes relevant to
  hypersonics. \emph{J.~Phys.\ Chem.~A} \textbf{2020}, \emph{124},
  6255--6269\relax
\mciteBstWouldAddEndPuncttrue
\mciteSetBstMidEndSepPunct{\mcitedefaultmidpunct}
{\mcitedefaultendpunct}{\mcitedefaultseppunct}\relax
\EndOfBibitem
\bibitem[Zanchet \latin{et~al.}(2009)Zanchet, Bussery-Honvault, Jorfi, and
  Honvault]{zanchet2009study}
Zanchet,~A.; Bussery-Honvault,~B.; Jorfi,~M.; Honvault,~P. Study of the
  C($^{3}$P)+ OH(X$^{2}\Pi$)$\rightarrow$ CO(a$^{3}\Pi$)+ H($^{2}$S) reaction:
  fully global ab initio potential energy surfaces of the 1$^{2}$A${"}$ and
  1$^{4}$A${"}$ excited states and non adiabatic couplings. \emph{Phys.\ Chem.\
  Chem.\ Phys.} \textbf{2009}, \emph{11}, 6182--6191\relax
\mciteBstWouldAddEndPuncttrue
\mciteSetBstMidEndSepPunct{\mcitedefaultmidpunct}
{\mcitedefaultendpunct}{\mcitedefaultseppunct}\relax
\EndOfBibitem
\bibitem[Goswami \latin{et~al.}(2014)Goswami, Rao, Mahapatra, Bussery-Honvault,
  and Honvault]{goswami2014time}
Goswami,~S.; Rao,~T.~R.; Mahapatra,~S.; Bussery-Honvault,~B.; Honvault,~P.
  Time-dependent quantum wave packet dynamics of S + OH reaction on its
  electronic ground state. \emph{J.~Phys.\ Chem.~A} \textbf{2014}, \emph{118},
  5915--5926\relax
\mciteBstWouldAddEndPuncttrue
\mciteSetBstMidEndSepPunct{\mcitedefaultmidpunct}
{\mcitedefaultendpunct}{\mcitedefaultseppunct}\relax
\EndOfBibitem
\bibitem[Jorfi \latin{et~al.}(2009)Jorfi, Honvault, and
  Halvick]{jorfi2009quasi}
Jorfi,~M.; Honvault,~P.; Halvick,~P. Quasi-classical determination of integral
  cross-sections and rate constants for the N+ OH $\rightarrow$ NO+ H reaction.
  \emph{Chem. Phys. Lett.} \textbf{2009}, \emph{471}, 65--70\relax
\mciteBstWouldAddEndPuncttrue
\mciteSetBstMidEndSepPunct{\mcitedefaultmidpunct}
{\mcitedefaultendpunct}{\mcitedefaultseppunct}\relax
\EndOfBibitem
\bibitem[Jorfi \latin{et~al.}(2009)Jorfi, Honvault, and
  Halvick]{jorfi2009quasiclassical}
Jorfi,~M.; Honvault,~P.; Halvick,~P. Quasiclassical trajectory calculations of
  differential cross sections and product energy distributions for the N+ OH
  $\rightarrow$ NO+ H reaction. \emph{J.~Chem.\ Phys.} \textbf{2009},
  \emph{131}, 094302\relax
\mciteBstWouldAddEndPuncttrue
\mciteSetBstMidEndSepPunct{\mcitedefaultmidpunct}
{\mcitedefaultendpunct}{\mcitedefaultseppunct}\relax
\EndOfBibitem
\bibitem[Jorfi \latin{et~al.}(2009)Jorfi, Honvault, Bargue{\~n}o,
  Gonz{\'a}lez-Lezana, Larr{\'e}garay, Bonnet, and
  Halvick]{jorfi2009statistical}
Jorfi,~M.; Honvault,~P.; Bargue{\~n}o,~P.; Gonz{\'a}lez-Lezana,~T.;
  Larr{\'e}garay,~P.; Bonnet,~L.; Halvick,~P. On the statistical behavior of
  the O+ OH $\rightarrow$ H+ O$_{2}$ reaction: A comparison between
  quasiclassical trajectory, quantum scattering, and statistical calculations.
  \emph{J.~Chem.\ Phys.} \textbf{2009}, \emph{130}, 184301\relax
\mciteBstWouldAddEndPuncttrue
\mciteSetBstMidEndSepPunct{\mcitedefaultmidpunct}
{\mcitedefaultendpunct}{\mcitedefaultseppunct}\relax
\EndOfBibitem
\bibitem[Jorfi and Honvault(2010)Jorfi, and Honvault]{jorfi2010quantum}
Jorfi,~M.; Honvault,~P. Quantum dynamics at the state-to-state level of the
  C+OH reaction on the first excited potential energy surface. \emph{J.~Phys.\
  Chem.~A} \textbf{2010}, \emph{114}, 4742--4747\relax
\mciteBstWouldAddEndPuncttrue
\mciteSetBstMidEndSepPunct{\mcitedefaultmidpunct}
{\mcitedefaultendpunct}{\mcitedefaultseppunct}\relax
\EndOfBibitem
\bibitem[Jorfi and Honvault(2011)Jorfi, and Honvault]{jorfi2011state}
Jorfi,~M.; Honvault,~P. State-to-state quantum dynamics calculations of the
  C+OH reaction on the second excited potential energy surface. \emph{J.~Phys.\
  Chem.~A} \textbf{2011}, \emph{115}, 8791--8796\relax
\mciteBstWouldAddEndPuncttrue
\mciteSetBstMidEndSepPunct{\mcitedefaultmidpunct}
{\mcitedefaultendpunct}{\mcitedefaultseppunct}\relax
\EndOfBibitem
\bibitem[Rao \latin{et~al.}(2013)Rao, Goswami, Mahapatra, Bussery-Honvault, and
  Honvault]{rao:2013}
Rao,~T.~R.; Goswami,~S.; Mahapatra,~S.; Bussery-Honvault,~B.; Honvault,~P.
  Time-dependent quantum wave packet dynamics of the C + OH reaction on the
  excited electronic state. \emph{J.~Chem.\ Phys.} \textbf{2013}, \emph{138},
  094318\relax
\mciteBstWouldAddEndPuncttrue
\mciteSetBstMidEndSepPunct{\mcitedefaultmidpunct}
{\mcitedefaultendpunct}{\mcitedefaultseppunct}\relax
\EndOfBibitem
\bibitem[Jorfi and Honvault(2011)Jorfi, and Honvault]{jorfi2011quasi}
Jorfi,~M.; Honvault,~P. Quasi-classical trajectory study of the S+OH
  $\rightarrow$ SO+H reaction: from reaction probability to thermal rate
  constant. \emph{Phys.\ Chem.\ Chem.\ Phys.} \textbf{2011}, \emph{13},
  8414--8421\relax
\mciteBstWouldAddEndPuncttrue
\mciteSetBstMidEndSepPunct{\mcitedefaultmidpunct}
{\mcitedefaultendpunct}{\mcitedefaultseppunct}\relax
\EndOfBibitem
\bibitem[Panda \latin{et~al.}(2012)Panda, Herr{\'a}ez-Aguilar, Jambrina,
  Aldegunde, Althorpe, and Aoiz]{panda2012state}
Panda,~A.~N.; Herr{\'a}ez-Aguilar,~D.; Jambrina,~P.~G.; Aldegunde,~J.;
  Althorpe,~S.~C.; Aoiz,~F.~J. A state-to-state dynamical study of the Br +
  H$_{2}$ reaction: comparison of quantum and classical trajectory results.
  \emph{Phys.\ Chem.\ Chem.\ Phys.} \textbf{2012}, \emph{14},
  13067--13075\relax
\mciteBstWouldAddEndPuncttrue
\mciteSetBstMidEndSepPunct{\mcitedefaultmidpunct}
{\mcitedefaultendpunct}{\mcitedefaultseppunct}\relax
\EndOfBibitem
\bibitem[Bargue{\~n}o \latin{et~al.}(2011)Bargue{\~n}o, Jambrina, Alvari{\~n}o,
  Men{\'e}ndez, Verdasco, Hankel, Smith, Aoiz, and
  Gonz{\'a}lez-Lezana]{bargueno2011energy}
Bargue{\~n}o,~P.; Jambrina,~P.; Alvari{\~n}o,~J.~M.; Men{\'e}ndez,~M.;
  Verdasco,~E.; Hankel,~M.; Smith,~S.; Aoiz,~F.~J.; Gonz{\'a}lez-Lezana,~T.
  Energy dependent dynamics of the O($^{1}$D) + HCl reaction: A quantum,
  quasiclassical and statistical study. \emph{Phys.\ Chem.\ Chem.\ Phys.}
  \textbf{2011}, \emph{13}, 8502--8514\relax
\mciteBstWouldAddEndPuncttrue
\mciteSetBstMidEndSepPunct{\mcitedefaultmidpunct}
{\mcitedefaultendpunct}{\mcitedefaultseppunct}\relax
\EndOfBibitem
\bibitem[Hankel and Yue(2012)Hankel, and Yue]{hankel2012quantum}
Hankel,~M.; Yue,~X.-F. Quantum dynamics study of the N($^{2}$D) + H$_{2}$
  reaction and the effects of the potential energy surface on the propagation
  time. \emph{Comput.\ Theor.\ Chem.} \textbf{2012}, \emph{990}, 23--29\relax
\mciteBstWouldAddEndPuncttrue
\mciteSetBstMidEndSepPunct{\mcitedefaultmidpunct}
{\mcitedefaultendpunct}{\mcitedefaultseppunct}\relax
\EndOfBibitem
\bibitem[Goswami \latin{et~al.}(2017)Goswami, Bussery-Honvault, Honvault, and
  Mahapatra]{GoswamiMolPhys2017}
Goswami,~S.; Bussery-Honvault,~B.; Honvault,~P.; Mahapatra,~S. Effect of
  internal excitations of reagent diatom on initial state-selected dynamics of
  C + OH reaction on its second excited (1$^{4}$A${''}$) electronic state.
  \emph{Mol. Phys.} \textbf{2017}, \emph{115}, 2658--2672\relax
\mciteBstWouldAddEndPuncttrue
\mciteSetBstMidEndSepPunct{\mcitedefaultmidpunct}
{\mcitedefaultendpunct}{\mcitedefaultseppunct}\relax
\EndOfBibitem
\bibitem[Goswami \latin{et~al.}(2018)Goswami, Sahoo, Rao, Bussery-Honvault,
  Honvault, and Mahapatra]{goswami2018theoretical}
Goswami,~S.; Sahoo,~J.; Rao,~T.~R.; Bussery-Honvault,~B.; Honvault,~P.;
  Mahapatra,~S. A theoretical study on the C + OH reaction dynamics and product
  energy disposal with vibrationally excited reagent. \emph{Eur.\ Phys.\ J.~D}
  \textbf{2018}, \emph{72}, 1--19\relax
\mciteBstWouldAddEndPuncttrue
\mciteSetBstMidEndSepPunct{\mcitedefaultmidpunct}
{\mcitedefaultendpunct}{\mcitedefaultseppunct}\relax
\EndOfBibitem
\bibitem[Polanyi(1972)]{polanyi1972some}
Polanyi,~J. Some concepts in reaction dynamics. \emph{Acc. Chem. Res.}
  \textbf{1972}, \emph{5}, 161\relax
\mciteBstWouldAddEndPuncttrue
\mciteSetBstMidEndSepPunct{\mcitedefaultmidpunct}
{\mcitedefaultendpunct}{\mcitedefaultseppunct}\relax
\EndOfBibitem
\bibitem[Jiang and Guo(2013)Jiang, and Guo]{jiang2013relative}
Jiang,~B.; Guo,~H. Relative efficacy of vibrational vs. translational
  excitation in promoting atom-diatom reactivity: Rigorous examination of
  Polanyi's rules and proposition of sudden vector projection (SVP) model.
  \emph{J.~Chem.\ Phys.} \textbf{2013}, \emph{138}, 234104\relax
\mciteBstWouldAddEndPuncttrue
\mciteSetBstMidEndSepPunct{\mcitedefaultmidpunct}
{\mcitedefaultendpunct}{\mcitedefaultseppunct}\relax
\EndOfBibitem
\bibitem[Polanyi(1987)]{polanyi1987some}
Polanyi,~J.~C. Some concepts in reaction dynamics. \emph{Science}
  \textbf{1987}, \emph{236}, 680--690\relax
\mciteBstWouldAddEndPuncttrue
\mciteSetBstMidEndSepPunct{\mcitedefaultmidpunct}
{\mcitedefaultendpunct}{\mcitedefaultseppunct}\relax
\EndOfBibitem
\bibitem[Bulut \latin{et~al.}(2011)Bulut, Roncero, Jorfi, and
  Honvault]{bulut2011accurate}
Bulut,~N.; Roncero,~O.; Jorfi,~M.; Honvault,~P. Accurate time dependent wave
  packet calculations for the N+OH reaction. \emph{J.~Chem.\ Phys.}
  \textbf{2011}, \emph{135}, 104307\relax
\mciteBstWouldAddEndPuncttrue
\mciteSetBstMidEndSepPunct{\mcitedefaultmidpunct}
{\mcitedefaultendpunct}{\mcitedefaultseppunct}\relax
\EndOfBibitem
\bibitem[Lin \latin{et~al.}(2008)Lin, Guo, Honvault, Xu, and
  Xie]{lin2008accurate}
Lin,~S.~Y.; Guo,~H.; Honvault,~P.; Xu,~C.; Xie,~D. Accurate quantum mechanical
  calculations of differential and integral cross sections and rate constant
  for the O+ OH reaction using an ab initio potential energy surface.
  \emph{J.~Chem.\ Phys.} \textbf{2008}, \emph{128}, 014303\relax
\mciteBstWouldAddEndPuncttrue
\mciteSetBstMidEndSepPunct{\mcitedefaultmidpunct}
{\mcitedefaultendpunct}{\mcitedefaultseppunct}\relax
\EndOfBibitem
\bibitem[Hankel \latin{et~al.}(2008)Hankel, Smith, Gray, and
  Balint-Kurti]{hankel2008diffrealwave}
Hankel,~M.; Smith,~S.~C.; Gray,~S.~K.; Balint-Kurti,~G.~G. DIFFREALWAVE: A
  parallel real wavepacket code for the quantum mechanical calculation of
  reactive state-to-state differential cross sections in atom plus diatom
  collisions. \emph{Comput.\ Phys.\ Commun.} \textbf{2008}, \emph{179},
  569--578\relax
\mciteBstWouldAddEndPuncttrue
\mciteSetBstMidEndSepPunct{\mcitedefaultmidpunct}
{\mcitedefaultendpunct}{\mcitedefaultseppunct}\relax
\EndOfBibitem
\bibitem[Gray and Balint-Kurti(1998)Gray, and Balint-Kurti]{gray1998quantum}
Gray,~S.~K.; Balint-Kurti,~G.~G. Quantum dynamics with real wave packets,
  including application to three-dimensional $(J=0)$ D+H$_{2}\rightarrow$HD+H
  reactive scattering. \emph{J.~Chem.\ Phys.} \textbf{1998}, \emph{108},
  950--962\relax
\mciteBstWouldAddEndPuncttrue
\mciteSetBstMidEndSepPunct{\mcitedefaultmidpunct}
{\mcitedefaultendpunct}{\mcitedefaultseppunct}\relax
\EndOfBibitem
\bibitem[Hankel \latin{et~al.}(2006)Hankel, Smith, Allan, Gray, and
  Balint-Kurti]{hankel2006state}
Hankel,~M.; Smith,~S.~C.; Allan,~R.~J.; Gray,~S.~K.; Balint-Kurti,~G.~G.
  State-to-state reactive differential cross sections for the H + H$_{2}
  \rightarrow$ H$_{2}$ + H reaction on five different potential energy surfaces
  employing a new quantum wavepacket computer code: DIFFREALWAVE.
  \emph{J.~Chem.\ Phys.} \textbf{2006}, \emph{125}, 164303\relax
\mciteBstWouldAddEndPuncttrue
\mciteSetBstMidEndSepPunct{\mcitedefaultmidpunct}
{\mcitedefaultendpunct}{\mcitedefaultseppunct}\relax
\EndOfBibitem
\bibitem[Goswami \latin{et~al.}(2020)Goswami, Sahoo, Paul, Rao, and
  Mahapatra]{goswami2020effect}
Goswami,~S.; Sahoo,~J.; Paul,~S.~K.; Rao,~T.~R.; Mahapatra,~S. Effect of
  Reagent Vibration and Rotation on the State-to-State Dynamics of the Hydrogen
  Exchange Reaction, H + H$_{2} \rightarrow$ H$_{2}$+ H. \emph{J.~Phys.\
  Chem.~A} \textbf{2020}, \emph{124}, 9343--9359\relax
\mciteBstWouldAddEndPuncttrue
\mciteSetBstMidEndSepPunct{\mcitedefaultmidpunct}
{\mcitedefaultendpunct}{\mcitedefaultseppunct}\relax
\EndOfBibitem
\bibitem[Hankel \latin{et~al.}(2003)Hankel, Balint-Kurti, and
  Gray]{hankel2003sinc}
Hankel,~M.; Balint-Kurti,~G.~G.; Gray,~S.~K. Sinc wave packets: New form of
  wave packet for time-dependent quantum mechanical reactive scattering
  calculations. \emph{Int. J Quantum Chem.} \textbf{2003}, \emph{92},
  205--211\relax
\mciteBstWouldAddEndPuncttrue
\mciteSetBstMidEndSepPunct{\mcitedefaultmidpunct}
{\mcitedefaultendpunct}{\mcitedefaultseppunct}\relax
\EndOfBibitem
\bibitem[Offer and Balint-Kurti(1994)Offer, and Balint-Kurti]{offer1994time}
Offer,~A.~R.; Balint-Kurti,~G.~G. Time-dependent quantum mechanical study of
  the photodissociation of HOCl and DOCl. \emph{J.~Chem.\ Phys.} \textbf{1994},
  \emph{101}, 10416--10428\relax
\mciteBstWouldAddEndPuncttrue
\mciteSetBstMidEndSepPunct{\mcitedefaultmidpunct}
{\mcitedefaultendpunct}{\mcitedefaultseppunct}\relax
\EndOfBibitem
\bibitem[Kosloff and Kosloff(1983)Kosloff, and Kosloff]{kosloff1983fourier}
Kosloff,~D.; Kosloff,~R. A Fourier method solution for the time dependent
  Schr{\"o}dinger equation as a tool in molecular dynamics. \emph{J. Comput.
  Phys.} \textbf{1983}, \emph{52}, 35--53\relax
\mciteBstWouldAddEndPuncttrue
\mciteSetBstMidEndSepPunct{\mcitedefaultmidpunct}
{\mcitedefaultendpunct}{\mcitedefaultseppunct}\relax
\EndOfBibitem
\bibitem[Light \latin{et~al.}(1985)Light, Hamilton, and
  Lill]{light1985generalized}
Light,~J.; Hamilton,~I.; Lill,~J. Generalized discrete variable approximation
  in quantum mechanics. \emph{J.~Chem.\ Phys.} \textbf{1985}, \emph{82},
  1400--1409\relax
\mciteBstWouldAddEndPuncttrue
\mciteSetBstMidEndSepPunct{\mcitedefaultmidpunct}
{\mcitedefaultendpunct}{\mcitedefaultseppunct}\relax
\EndOfBibitem
\bibitem[Lill \latin{et~al.}(1982)Lill, Parker, and Light]{lill1982discrete}
Lill,~J.; Parker,~G.; Light,~J. Discrete variable representations and sudden
  models in quantum scattering theory. \emph{Chem.\ Phys.\ Lett.}
  \textbf{1982}, \emph{89}, 483--489\relax
\mciteBstWouldAddEndPuncttrue
\mciteSetBstMidEndSepPunct{\mcitedefaultmidpunct}
{\mcitedefaultendpunct}{\mcitedefaultseppunct}\relax
\EndOfBibitem
\bibitem[Hamilton and Light(1986)Hamilton, and Light]{hamilton1986distributed}
Hamilton,~I.; Light,~J. On distributed Gaussian bases for simple model
  multidimensional vibrational problems. \emph{J.~Chem.\ Phys.} \textbf{1986},
  \emph{84}, 306--317\relax
\mciteBstWouldAddEndPuncttrue
\mciteSetBstMidEndSepPunct{\mcitedefaultmidpunct}
{\mcitedefaultendpunct}{\mcitedefaultseppunct}\relax
\EndOfBibitem
\bibitem[Henriksen and Hansen(2011)Henriksen, and Hansen]{hen11}
Henriksen,~N.~E.; Hansen,~F.~Y. \emph{Theories of Molecular Reaction Dynamics};
  Oxford, 2011\relax
\mciteBstWouldAddEndPuncttrue
\mciteSetBstMidEndSepPunct{\mcitedefaultmidpunct}
{\mcitedefaultendpunct}{\mcitedefaultseppunct}\relax
\EndOfBibitem
\bibitem[Koner \latin{et~al.}(2016)Koner, Barrios, Gonz\'{a}lez-Lezana, and
  Panda]{kon16:4731}
Koner,~D.; Barrios,~L.; Gonz\'{a}lez-Lezana,~T.; Panda,~A.~N. State-to-State
  Dynamics of the Ne + HeH$^+ (v = 0, j = 0) \rightarrow$ NeH$^+(v', j')$ + He
  Reaction. \emph{J.~Phys.\ Chem.~A} \textbf{2016}, \emph{120},
  4731--4741\relax
\mciteBstWouldAddEndPuncttrue
\mciteSetBstMidEndSepPunct{\mcitedefaultmidpunct}
{\mcitedefaultendpunct}{\mcitedefaultseppunct}\relax
\EndOfBibitem
\bibitem[Koner \latin{et~al.}(2018)Koner, Bemish, and Meuwly]{MM.cno:2018}
Koner,~D.; Bemish,~R.~J.; Meuwly,~M. The C($^{3}$P) + NO(X$^{2} \Pi$)
  $\rightarrow$ O($^{3}$P) + CN(X$^{2}\Sigma^{+}$), N($^{2}$D)/N($^{4}$S) +
  CO(X$^{1}\Sigma^{+}$) reaction: Rates, branching ratios, and final states
  from 15 K to 20 000 K. \emph{J.~Chem.\ Phys.} \textbf{2018}, \emph{149},
  094305\relax
\mciteBstWouldAddEndPuncttrue
\mciteSetBstMidEndSepPunct{\mcitedefaultmidpunct}
{\mcitedefaultendpunct}{\mcitedefaultseppunct}\relax
\EndOfBibitem
\bibitem[Truhlar and Muckerman(1979)Truhlar, and Muckerman]{tru79}
Truhlar,~D.~G.; Muckerman,~J.~T. In \emph{Atom - Molecule Collision Theory};
  Bernstein,~R.~B., Ed.; Springer US, 1979; pp 505--566\relax
\mciteBstWouldAddEndPuncttrue
\mciteSetBstMidEndSepPunct{\mcitedefaultmidpunct}
{\mcitedefaultendpunct}{\mcitedefaultseppunct}\relax
\EndOfBibitem
\bibitem[Bonnet and Rayez(1997)Bonnet, and Rayez]{bon97:183}
Bonnet,~L.; Rayez,~J.-C. Quasiclassical Trajectory Method for Molecular
  Scattering Processes: Necessity of a Weighted Binning Approach. \emph{Chem.
  Phys. Lett.} \textbf{1997}, \emph{277}, 183--190\relax
\mciteBstWouldAddEndPuncttrue
\mciteSetBstMidEndSepPunct{\mcitedefaultmidpunct}
{\mcitedefaultendpunct}{\mcitedefaultseppunct}\relax
\EndOfBibitem
\bibitem[Bonnet and Rayez(2004)Bonnet, and Rayez]{bon04:106}
Bonnet,~L.; Rayez,~J.-C. Gaussian Weighting in the Quasiclassical Trajectory
  Method. \emph{Chem. Phys. Lett.} \textbf{2004}, \emph{397}, 106--109\relax
\mciteBstWouldAddEndPuncttrue
\mciteSetBstMidEndSepPunct{\mcitedefaultmidpunct}
{\mcitedefaultendpunct}{\mcitedefaultseppunct}\relax
\EndOfBibitem
\bibitem[Jorfi and Honvault(2009)Jorfi, and Honvault]{jorfi2009state}
Jorfi,~M.; Honvault,~P. State-to-state quantum dynamical study of the N + OH
  $\rightarrow$ NO + H reaction. \emph{J.~Phys.\ Chem.~A} \textbf{2009},
  \emph{113}, 2316--2322\relax
\mciteBstWouldAddEndPuncttrue
\mciteSetBstMidEndSepPunct{\mcitedefaultmidpunct}
{\mcitedefaultendpunct}{\mcitedefaultseppunct}\relax
\EndOfBibitem
\bibitem[Ba{\~n}ares \latin{et~al.}(2003)Ba{\~n}ares, Aoiz, Honvault,
  Bussery-Honvault, and Launay]{banares2003quantum}
Ba{\~n}ares,~L.; Aoiz,~F.; Honvault,~P.; Bussery-Honvault,~B.; Launay,~J.-M.
  Quantum mechanical and quasi-classical trajectory study of the C($^{1}$D) +
  H$_{2}$ reaction dynamics. \emph{J.~Chem.\ Phys.} \textbf{2003}, \emph{118},
  565--568\relax
\mciteBstWouldAddEndPuncttrue
\mciteSetBstMidEndSepPunct{\mcitedefaultmidpunct}
{\mcitedefaultendpunct}{\mcitedefaultseppunct}\relax
\EndOfBibitem
\bibitem[Banares \latin{et~al.}(2005)Banares, Castillo, Honvault, and
  Launay]{banares2005quantum}
Banares,~L.; Castillo,~J.; Honvault,~P.; Launay,~J.-M. Quantum mechanical and
  quasi-classical trajectory reaction probabilities and cross sections for the
  S($^{1}$D) + H$_{2}$, D$_{2}$, HD insertion reactions. \emph{Phys.\ Chem.\
  Chem.\ Phys.} \textbf{2005}, \emph{7}, 627--634\relax
\mciteBstWouldAddEndPuncttrue
\mciteSetBstMidEndSepPunct{\mcitedefaultmidpunct}
{\mcitedefaultendpunct}{\mcitedefaultseppunct}\relax
\EndOfBibitem
\bibitem[Bulut \latin{et~al.}(2009)Bulut, Zanchet, Honvault, Bussery-Honvault,
  and Ba{\~n}ares]{bulut2009time}
Bulut,~N.; Zanchet,~A.; Honvault,~P.; Bussery-Honvault,~B.; Ba{\~n}ares,~L.
  Time-dependent wave packet and quasiclassical trajectory study of the
  C($^{3}$P)+OH(X$^{2}\Pi$)$\rightarrow$ CO(X$^{1}\Sigma^{+}$)+H($^{2}$S)
  reaction at the state-to-state level. \emph{J.~Chem.\ Phys.} \textbf{2009},
  \emph{130}, 194303\relax
\mciteBstWouldAddEndPuncttrue
\mciteSetBstMidEndSepPunct{\mcitedefaultmidpunct}
{\mcitedefaultendpunct}{\mcitedefaultseppunct}\relax
\EndOfBibitem
\bibitem[Upadhyay \latin{et~al.}(2021)Upadhyay, Pezzella, and
  Meuwly]{MM.co2:2021}
Upadhyay,~M.; Pezzella,~M.; Meuwly,~M. Genesis of Polyatomic Molecules in Dark
  Clouds: CO2 Formation on Cold Amorphous Solid Water. \emph{J.~Phys.\ Chem.
  Lett.} \textbf{2021}, \emph{12}, 6781--6787\relax
\mciteBstWouldAddEndPuncttrue
\mciteSetBstMidEndSepPunct{\mcitedefaultmidpunct}
{\mcitedefaultendpunct}{\mcitedefaultseppunct}\relax
\EndOfBibitem
\bibitem[Upadhyay and Meuwly(2021)Upadhyay, and Meuwly]{MM.co2.2:2021}
Upadhyay,~M.; Meuwly,~M. Energy Redistribution Following CO2 Formation on Cold
  Amorphous Solid Water. \emph{Frontiers in chemistry} \textbf{2021},
  \emph{9}\relax
\mciteBstWouldAddEndPuncttrue
\mciteSetBstMidEndSepPunct{\mcitedefaultmidpunct}
{\mcitedefaultendpunct}{\mcitedefaultseppunct}\relax
\EndOfBibitem
\bibitem[Pezzella \latin{et~al.}(2018)Pezzella, Unke, and Meuwly]{MM.o2:2018}
Pezzella,~M.; Unke,~O.~T.; Meuwly,~M. Molecular Oxygen Formation in
  Interstellar Ices Does Not Require Tunneling. \emph{J.~Phys.\ Chem. Lett.}
  \textbf{2018}, \emph{9}, 1822--1826\relax
\mciteBstWouldAddEndPuncttrue
\mciteSetBstMidEndSepPunct{\mcitedefaultmidpunct}
{\mcitedefaultendpunct}{\mcitedefaultseppunct}\relax
\EndOfBibitem
\bibitem[Pezzella \latin{et~al.}(2020)Pezzella, Koner, and Meuwly]{MM.o2:2020}
Pezzella,~M.; Koner,~D.; Meuwly,~M. Formation and Stabilization of Ground and
  Excited-State Singlet O$_2$ upon Recombination of $^3$P Oxygen on Amorphous
  Solid Water. \emph{J.~Phys.\ Chem. Lett.} \textbf{2020}, \emph{11},
  2171--2176\relax
\mciteBstWouldAddEndPuncttrue
\mciteSetBstMidEndSepPunct{\mcitedefaultmidpunct}
{\mcitedefaultendpunct}{\mcitedefaultseppunct}\relax
\EndOfBibitem
\bibitem[Costes and Naulin(1998)Costes, and Naulin]{costes:1998}
Costes,~M.; Naulin,~C. State-to-state cross sections for the C($^3$P$_J$)+
  O$_2$(X$^3 \Sigma_{g}^{-}$) $\rightarrow$ CO(X$^1 \Sigma^{+}$)+ O($^1$D$_2$)
  reaction at kinetic energies between 4.4 and 90 meV. \emph{Comptes Rendus de
  l{'}Acad{\'e}mie des Sciences-Series IIC-Chemistry} \textbf{1998}, \emph{1},
  771--775\relax
\mciteBstWouldAddEndPuncttrue
\mciteSetBstMidEndSepPunct{\mcitedefaultmidpunct}
{\mcitedefaultendpunct}{\mcitedefaultseppunct}\relax
\EndOfBibitem
\end{mcitethebibliography}

\end{document}